\documentclass[12pt]{article}

\usepackage{amsmath,amssymb,amsfonts}	
\usepackage{cite}
\usepackage{color}
\usepackage{amsmath}
\usepackage{amsfonts}
\usepackage{amssymb}
\usepackage{graphicx}
\usepackage{slashed}            
\usepackage{ulem}

\usepackage{tikz}
\usetikzlibrary{arrows,shapes}
\usetikzlibrary{trees}
\usetikzlibrary{matrix,arrows} 				
\usetikzlibrary{positioning}				
\usetikzlibrary{calc,through}				
\usetikzlibrary{decorations.pathreplacing}  
\usepackage{pgffor}							

\usetikzlibrary{decorations.pathmorphing}	
\usetikzlibrary{decorations.markings}
\tikzset{
    vector/.style={decorate, decoration={snake}, draw},
	provector/.style={decorate, decoration={snake,amplitude=2.5pt}, draw},
	antivector/.style={decorate, decoration={snake,amplitude=-2.5pt}, draw},
    fermion/.style={draw=black, postaction={decorate},
        decoration={markings,mark=at position .55 with {\arrow[draw=black]{>}}}},
    fermionbar/.style={draw=black, postaction={decorate},
        decoration={markings,mark=at position .55 with {\arrow[draw=black]{<}}}},
    fermionnoarrow/.style={draw=black},
    gluon/.style={decorate, draw=black,
        decoration={coil,amplitude=4pt, segment length=5pt}},
    scalar/.style={dashed,draw=black, postaction={decorate},
        decoration={markings,mark=at position .55 with {\arrow[draw=black]{>}}}},
    scalarbar/.style={dashed,draw=black, postaction={decorate},
        decoration={markings,mark=at position .55 with {\arrow[draw=black]{<}}}},
    scalarnoarrow/.style={dashed,draw=black},
    electron/.style={draw=black, postaction={decorate},
        decoration={markings,mark=at position .55 with {\arrow[draw=black]{>}}}},
	bigvector/.style={decorate, decoration={snake,amplitude=4pt}, draw},
}

\tikzstyle{block} = [draw, rectangle, 
    minimum height=3em, minimum width=6em]

\usepackage{hyperref}			

\def\be{\begin{equation}}
\def\ee{\end{equation}}
\def\bea{\begin{eqnarray}}
\def\eea{\end{eqnarray}}

\catcode`\@=11
\def\lsim{\mathrel{\mathpalette\@versim<}}
\def\gsim{\mathrel{\mathpalette\@versim>}}
\def\@versim#1#2{\vcenter{\offinterlineskip
\ialign{$\m@th#1\hfil##\hfil$\crcr#2\crcr\sim\crcr } }}
\catcode`\@=12
\usepackage{axodraw}

\catcode`@=12
\begin{document}
\thispagestyle{empty}
\begin{flushright}
UCRHEP-533\\
August 2013\
\end{flushright}
\vspace{0.3in}
\begin{center}
{\LARGE \bf Supersymmetric Left-Right Model of\\ 
Radiative Neutrino Mass with\\ Multipartite Dark Matter\\}
\vspace{1.2in}
{\bf Subhaditya Bhattacharya \footnote{E-mail: subhaditya123@gmail.com}, Ernest Ma \footnote{E-mail: ma@phyun8.ucr.edu}, and Daniel Wegman \footnote{E-mail: wegman.daniel@gmail.com}\\}
\vspace{0.2in}
{\sl Department of Physics and Astronomy, University of California,\\ 
Riverside, California 92521, USA\\}
\end{center}
\vspace{0.6in}
\begin{abstract}\
The unifiable supersymmetric left-right model where the neutral fermion $n$ 
in the $SU(2)_R$ doublet $(n,e)_R$ is a dark-matter candidate, is shown to 
have the requisite particle content for the neutrino $\nu$ in the 
$SU(2)_L$ doublet $(\nu,e)_L$  to acquire a small radiative Majorana mass
from dark matter, i.e. scotogenic from the Greek ``scotos'' meaning 
darkness.  As a result, there are at least three coexisting stable
dark-matter particles with different interactions.  We study their 
possible phenomenological impact on present and future experiments.
\end{abstract}

\newpage
\baselineskip 24pt

\section{Introduction}

To understand dark matter in the context of extensions of the standard 
model of particle interactions, there are many avenues.  Supersymmetry 
with $R-$parity conservation is the most common approach.  Two other 
well-motivated scenarios have also been proposed in recent years. 
One is the idea of radiative neutrino mass induced by dark matter. 
The simplest such one-loop mechanism was proposed by one of us in 
2006~\cite{m06}.  It has been called ``scotogenic'' from the Greek 
``scotos'' meaning darkness.  This proposal has been studied and 
extended in a number of subsequent papers~\cite{kms06,m06-1,m06-2,ms07,
hkmr07,m08,bm08,m08-3,m08-1,m08-2,ms09,m09-1,m09-2,fm12,mnr12,m12-1}.  
Another is to have a left-right extension where the neutral component $n$ of 
the $SU(2)_R$ doublet $(n,e)_R$ is dark matter~\cite{klm09,m09,adhm10,
klm10,m10,m12}.  

It was pointed out~\cite{m10} that with the addition of new supermultiplets, 
the dark left-right model is unifiable with all gauge couplings converging 
at an energy scale of about $10^{16}$ GeV.  These additional particles 
turn out to be exactly what are required for radiative neutrino masses 
in the scotogenic model~\cite{m06}.  Hence an opportunity exists for 
merging all three mechanisms for dark matter in the context of a 
supersymmetric unified theory of radiative neutrino masses.  In this paper 
we will focus mainly on the dark matter (DM) phenomenology of this
comprehensive model.

\section{Model}
Consider the gauge group $SU(3)_C \times SU(2)_L \times SU(2)_R \times U(1)$.  
A new global $U(1)$ symmetry $S$ is imposed so that the spontaneous breaking 
of $SU(2)_R \times S$ will leave the combination $S' = S + T_{3R}$ unbroken. 
Under $SU(3)_C \times SU(2)_L \times SU(2)_R \times U(1) \times S \times M 
\times H$, where $M$ and $H$ are discrete $Z_2$ symmetries, with the usual 
$R$ parity of the MSSM given by $R \equiv MH(-1)^{2j}$, the superfields 
transform as shown in Table 1.  
Because of supersymmetry, the Higgs sector is doubled, in analogy to the 
transition from the Standard Model (SM) to the Minimal Supersymmetric 
Standard Model (MSSM).  Another set of Higgs doublet superfields $\eta$ 
and a new set of charged and neutral Higgs singlet superfields $\zeta$ are 
added to obtain gauge-coupling unification, as well as radiative seesaw 
neutrino masses.

\begin{table}[htb]
\begin{center}
\begin{tabular}{|c|c|c|c|c|}
\hline
Superfield & $SU(3)_C \times SU(2)_L \times SU(2)_R \times U(1)$ & $S$ & $M$ 
& $H$ \\
\hline
$\psi = (\nu,e)$ & $(1,2,1,-1/2)$ & $0$ & $-$ & + \\
$\psi^c = (e^c,n^c)$ & $(1,1,2,1/2)$ & $-1/2$ & $-$ & + \\
$N$ & $(1,1,1,0)$ & $0$ & $-$ & $-$ \\ 
$n$ & $(1,1,1,0)$ & $1$ & $-$ & + \\ 
\hline
$Q = (u,d)$ & $(3,2,1,1/6)$ & $0$ & $-$ & + \\
$Q^c = (h^c,u^c)$ & $(3^*,1,2,-1/6)$ & $1/2$ & $-$ & + \\
$d^c$ & $(3^*,1,1,1/3)$ & $0$ & $-$ & + \\
$h$ & $(3,1,1,-1/3)$ & $-1$ & $-$ & + \\
\hline
$\Delta_1$ & $(1,2,2,0)$ & $1/2$ & + & + \\
$\Delta_2$ & $(1,2,2,0)$ & $-1/2$ & + & + \\
\hline
$\Phi_{L1}$ & $(1,2,1,-1/2)$ & $0$ & + & + \\
$\Phi_{L2}$ & $(1,2,1,1/2)$ & $0$ & + & + \\
$\Phi_{R1}$ & $(1,1,2,-1/2)$ & $-1/2$ & + & + \\
$\Phi_{R2}$ & $(1,1,2,1/2)$ & $1/2$ & + & + \\
\hline
$\eta_{L1}$ & $(1,2,1,-1/2)$ & $0$ & + & $-$ \\
$\eta_{L2}$ & $(1,2,1,1/2)$ & $0$ & + & $-$ \\
$\eta_{R1}$ & $(1,1,2,-1/2)$ & $1/2$ & + & $-$ \\
$\eta_{R2}$ & $(1,1,2,1/2)$ & $-1/2$ & + & $-$ \\
\hline
$\zeta_1$ & $(1,1,1,-1)$ & $0$ & + & $-$ \\
$\zeta_2$ & $(1,1,1,1)$ & $0$ & + & $-$ \\
$\zeta_3$ & $(1,1,1,0)$ & $0$ & + & $-$ \\
\hline
\end{tabular}
\caption{Particle content of proposed model.}
\end{center}
\end{table}

The superpotential of the model reads:

\begin{eqnarray}
W&=&- \mu_L \Phi_{L1} \Phi_{L2} - \mu_R \Phi_{R1} \Phi_{R2} - \mu_{\Delta} Tr( \Delta_{1} \Delta_{2} )
 \\ \nonumber &-& \mu_{L2} \eta_{L1} \eta_{L2} - \mu_{R2} \eta_{R1} \eta_{R2} - \mu_{s12} \zeta_{1} \zeta_{2} - \mu_{s 3} \zeta_{3} \zeta_{3} \\ \nonumber &+& f_1 \Phi_{L1} \Delta_{2} \Phi_{R2} + f_2 \Phi_{L2} \Delta_{1} \Phi_{R1} \\ \nonumber
 &+& f_3 \eta_{L1} \Delta_{1} \eta_{R2} + f_4 \eta_{L2} \Delta_{2} \eta_{R1} +f_5 \Phi_{L1} \eta_{L1} \zeta_2 \\ \nonumber &+&f_6 \Phi_{R1} \eta_{R1} \zeta_2 + f_7 \Phi_{L2} \eta_{L2} \zeta_1 +f_8 \Phi_{R2} \eta_{R2} \zeta_1 \\ \nonumber &+& f_9 \Phi_{L1} \eta_{L2} \chi_3 +f_{10} \Phi_{L2} \eta_{L1} \zeta_3 \\ \nonumber
 &+& f_{11} \psi \Delta_1 \psi^c + f_{12} Q \Delta_2 Q^c +f_{13} Q \Phi_{L1} d^c \\ \nonumber &+& f_{14}n \psi^c \Phi_{R1} +f_{15} h Q^c \Phi_{R2} \\ \nonumber &+& f_{16} \psi N \eta_{L2} + f_{17} \psi^c N \eta_{R1}
  \end{eqnarray}

The symmetry $S \times M \times H$ is used here to distinguish 
$\psi$, $\Phi_{L1}$, $\eta_{L1}$ from one another, as well as 
$\psi^c$, $\Phi_{R2}$, $\eta_{R2}$, and $N, n, \zeta_3$.  
There are seven bilinear terms with coefficients $\mu$ and seventeen 
trilinear terms with coefficient $f$ allowed by $S \times M \times H$.


Hence $m_e$ comes from the $I_{3L} = 1/2$ and $I_{3R} = -1/2$ component of 
$\Delta_1$, i.e. $\delta^0_{11}$  ($\langle \delta^0_{11} \rangle = u_1$)
with $S'=1/2-1/2=0$, $m_u$ from the 
$I_{3L} = -1/2$ and $I_{3R} = 1/2$ component of $\Delta_2$, i.e. 
$\delta^0_{22}$  ($\langle \delta^0_{22} \rangle = u_4$)
with $S'=-1/2+1/2=0$, $m_d$ from $\phi^0_{L1}$ ($\langle \phi^0_{L1} \rangle = {v_L}_1$), 
$m_n$ from $\phi^0_{R1}$ ($\langle \phi^0_{R1} \rangle = {v_R}_1$), 
and $m_h$ from $\phi^0_{R2}$ ($\langle \phi^0_{R2} \rangle = {v_R}_2$).  
Note that $\phi^0_{L2} $ ($\langle \phi^0_{L2} \rangle = {v_L}_2$) doesn't contribute to fermion masses, 
but is involved in the scalar and vector masses. This structure 
guarantees the absence of tree-level flavor-changing neutral currents.

\section{Radiative seesaw neutrino masses}
Since the neutrino $\nu$ does not couple to $N$ through $\Phi_{L2}$, it 
has no tree-level mass.  However, the $\nu N \eta_{L2}^0$ and $\phi_{L1}^0 
\eta_{L2}^0 \zeta_3$ couplings and the allowed Majorana masses for $N$ 
and $\zeta_3$ will generate one-loop radiative seesaw neutrino masses, 
as shown in Fig.~1.

\begin{figure}
\vspace{1em}
\begin{tikzpicture}[line width=1.5 pt, scale=2.7]
	\draw[fermion] (0.2,0)--(.8,0);
	\draw[fermion] (2,.5) arc (45:0:.7);
	\draw[fermion] (2,.5) arc (45:90:.7);
	\draw[fermionbar] (1.49,.7) arc (90:135:.7);
	\draw[fermion] (1,.5) arc (135:180:.7);
	\draw[fermionbar,dashed] (.8,0) -- (1.5,0);
	\draw[fermion,dashed] (2,.5) -- (2.3,.8);
	\draw[fermionbar,dashed] (.7,.8) -- (1,.5);
	\draw[fermion,dashed] (1.5,0) -- (2.2,0);
	\draw[fermionbar] (2.2,0) --(2.8,0);
	\node at (1.5,0) {$X$};
	\node at (1.5,.7) {$X$};
	\node at (.5,-.12) {$\nu$};
	\node at (2.5,-.12) {$\nu$};
	\node at (1.2,-.12) {$\tilde{N}$};
	\node at (1.8,-.12) {$\tilde{N}$};
	\node at (.7,.3) {$\tilde{\eta}_{L2}$};
	\node at (2.3,.3) {$\tilde{\eta}_{L2}$};
	\node at (1.85,.75) {$\zeta_3$};
	\node at (1.1,.75) {$\zeta_3$};
	\node at (.63,.9) {$\phi_{L1}^0$};
	\node at (2.35,.9) {$\phi_{L1}^0$};
	
\begin{scope}[shift={(3,0)}]
	\draw[fermion] (0.2,0)--(1,0);
	\draw[fermion,dashed] (2,.5) arc (45:0:.7);
	\draw[fermion,dashed] (2,.5) arc (45:90:.7);
	\draw[fermionbar,dashed] (1.49,.7) arc (90:135:.7);
	\draw[fermion,dashed] (1,.5) arc (135:180:.7);
	\draw[fermionbar] (1,0) -- (1.5,0);
	\draw[fermion,dashed] (2,.5) -- (2.3,.8);
	\draw[fermionbar,dashed] (.7,.8) -- (1,.5);
	\draw[fermion] (1.5,0) -- (2,0);
	\node at (1.5,0) {$X$};
	\node at (1.5,.7) {$X$};
	\node at (.5,-.12) {$\nu$};
	\node at (2.5,-.12) {$\nu$};
	\node at (1.2,-.12) {$N$};
	\node at (1.8,-.12) {$N$};
	\node at (.7,.3) {$\eta_{L2}$};
	\node at (2.3,.3) {$\eta_{L2}$};
	\node at (1.85,.75) {$\tilde{\zeta}_3$};
	\node at (1.1,.75) {$\tilde{\zeta}_3$};
	\node at (.63,.9) {$\phi_{L1}^0$};
	\node at (2.35,.9) {$\phi_{L1}^0$};
	\draw[fermionbar] (2,0) --(2.8,0);
\end{scope}
\end{tikzpicture}
\caption{Scotogenic neutrino mass}
\end{figure}
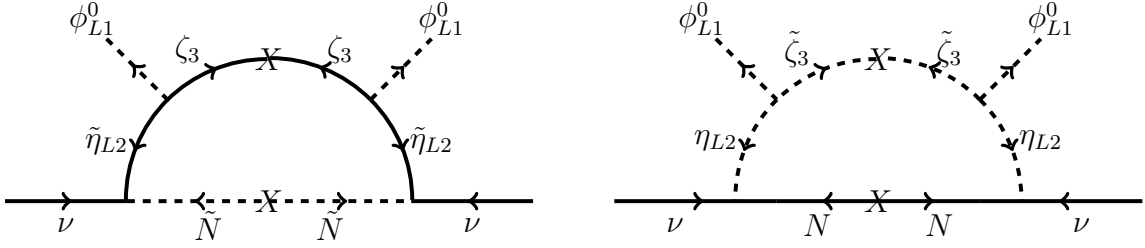

The loop can be calculated exactly via 
 
\begin{eqnarray}
(M_{\nu})_{\alpha\beta}=\sum_{i}\frac{h_{\alpha i}h_{\beta i}M_{Ni}}{16\pi^{2}} \bigg[ \sum_{j} (U_R)_{1j} \bigg[ \frac{m_{Rj}^{2}}{m_{Rj}^{2}-M_{Ni}^{2}}\ln\left(\frac{m_{Rj}^{2}}{M_{Ni}^{2}}\right) \bigg] \\ \nonumber
-  \sum_{j} (U_I)_{1j} \bigg[\frac{m_{Ij}^{2}}{m_{Ij}^{2}-M_{Ni}^{2}}\ln\left(\frac{m_{Ij}^{2}}{M_{Ni}^{2}}\right)\bigg] \bigg]
\end{eqnarray}
 
Where  $U_R$ ($U_I$) is the unitary matrix that makes $m_R$($m_I$) mass eigenstates and $h_{\alpha i}$ is the parameter for the interactions $\nu N \eta_{L2}^0$ and $\nu \tilde{N} \tilde{\eta_{L2}}^0$.

Both diagrams require supersymmetry breaking to be nonzero.  The one on the 
right needs the $A$ term $\phi^0_{L1} \eta^0_{L2} \tilde{\zeta}_3$ twice 
and the $B$ term $\tilde{\zeta}_3 \tilde{\zeta}_3$ once whereas the one on the 
left requires only the $B$ term $\tilde{N} \tilde{N}$ once.  We expect thus 
the latter diagram to be much more important.  We estimate its contribution 
to be given by
\begin{equation}
m_\nu \simeq {h^2 v_{L1}^2 \Delta M_N^2 \over 16 \pi^2 M_N^3},
\end{equation}
where $h$ is the diagonal Yukawa coupling, 
$\Delta M_N^2$ is the supersymmetry breaking $B$ term, and $M_3 \simeq M_N$ 
has been assumed.  Using $v_{L1} \simeq 100$ GeV, $\Delta M_N^2 \simeq 1$ TeV$^2$, 
and $M_N \simeq 10^5$ GeV, and $h^2 \simeq 10^{-3}$, we find $m_\nu \simeq 
0.1$ eV.

\section{Dark matter candidates of the model}
There are three conserved quantities in this model: a global $U(1)$ number 
$S' = S + T_{3R}$, with $S' = 0$ for the usual quarks and leptons and 
$S'=1$ for the scotino $n$, and the discrete $Z_2$ symmetries $M$ and $H$. 
The usual $R$ parity is then $R \equiv MH(-1)^{2j}$.  The various superfields 
of this model under $S'$, $M$, and $H$ are listed in Table 2.  
\begin{table}[htb]
\begin{center}
\begin{tabular}{|c|c|c|c|}
\hline
$S'$ & $M$ & $H$ & Superfields\\
\hline
0 & $-$ & + & $u,d,\nu,e$ \\ 
0 & + & + & $g,\gamma,W_L^\pm,Z,Z'$ \\
0 & + & + & $\phi^0_{L1},\phi^-_{L1},\phi^+_{L2},\phi^0_{L2},\phi^0_{R1},
\phi^0_{R2}$ \\
0 & + & + & $\delta^0_{11},\delta^-_{11},\delta^+_{22},\delta^0_{22}$ \\
\hline
1 & $-$ & + & $n,h^c$ \\ 
$-1$ & $-$ & + & $n^c,h$ \\ 
1 & + & + & $W_R^+,\phi^+_{R2},\delta^+_{12},\delta^0_{12}$ \\ 
$-1$ & + & + & $W_R^-,\phi^-_{R1},\delta^0_{21},\delta^-_{21}$ \\ 
\hline
0 & $-$ & $-$ & $N$ \\
$0$ & + & $-$ & $\eta^0_{L1},\eta^-_{L1},\eta^+_{L2},\eta^0_{L2},\eta^-_{R1},
\eta^+_{R2}$ \\ 
0 & + & $-$ & $\zeta^-_1,\zeta^+_2,\zeta_3$ \\ 
1 & + & $-$ & $\eta^0_{R1}$ \\ 
$-1$ & + & $-$ & $\eta^0_{R2}$ \\ 
\hline
\end{tabular}
\caption{Superfields under $S'=S+T_{3R}$, $M$, and $H$.}
\end{center}
\end{table}
A possible 
scenario for dark matter is to have the following three coexisting stable 
particles~\cite{cmwy07}: the lightest neutralino $\tilde{\chi}^0_1$ 
($S'=0$, $H=+$, $R=-$), the lightest scotino $n$ ($S'=1$, $H=+$, $R=+$), 
and the exotic $\tilde{\eta}^0_{R}$ fermion ($S'=1$, $H=-$, $R=+$). One should note here that 
the $\tilde{\eta}^0_{R}$ fermion is a type of neutralino but, it doesn't mix with gauginos 
and other Higgsinos and that's how it 
differs from the lightest neutralino $\tilde{\chi}^0_1$  of this model, 
making it possible to be much heavier than the LSP and still be stable. However, 
there may be additional stable particles due to kinematics.  For example, if 
the scalar counterpart of $n$ cannot 
decay into $n$ plus the lightest neutralino, then it will also be stable. 
There may even be several exotic stable $\eta$ and $\zeta$ particles.  The 
dark sector may be far from just the one particle that is usually assumed, 
as in the MSSM.  In the presence of several dark-matter candidates, the one 
with the largest annihilation cross section contributes the least, but may 
be the first to be discovered at the Large Hadron Collider (LHC).  This means 
that in this model, the severe constraint due to dark-matter relic abundance 
on the one candidate particle of the MSSM, i.e. the lightest neutralino, 
may be relaxed, because it needs only to account for a fraction of the 
total dark-matter abundance.  The allowed parameter space of the MSSM 
becomes much bigger and the opportunity for its discovery is enhanced 
at the LHC. 

\begin{figure}[thb]
$$
\includegraphics[height=6.5cm]{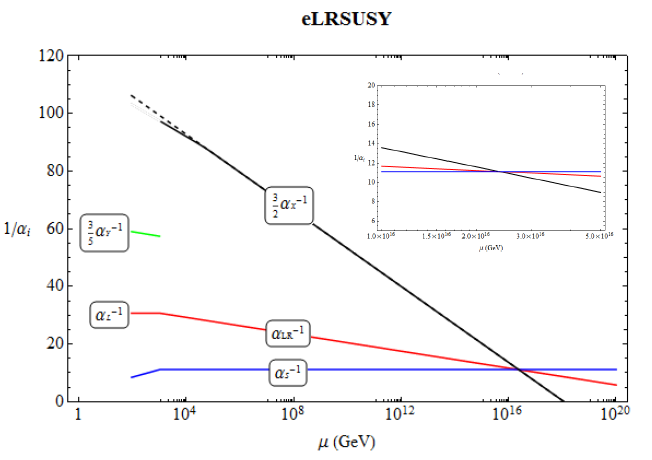}
$$
\caption{Gauge coupling unification in this model.}
\label{fig:unify}
\end{figure}

On Fig. \ref{fig:unify}, we show the gauge coupling unification of this model~\cite{m10}. 
The $U(1)_Y$ coupling runs until the $SU(2)_R$ breaking scale $M_R$, where 
$\alpha_X^{-1}(M_R)=\alpha_Y^{-1}(M_R)+\alpha_L^{-1}(M_R)$ and 
$\alpha_L=\alpha_R$. After $M_R$, the gauge symmetry $SU(2)_L$ and 
$SU(2)_R$ are unified with a coupling $\alpha_{LR}$. 
From the requirement of gauge-coupling unification, it was shown 
that if the $SU(2)_R$ breaking scale $M_R$ equals to the supersymmetry 
breaking scale $M_S$, the mass scale $M_X$ of the singlet superfields 
$\zeta_{1,2,3}$ should obey
\begin{equation}
M_R^{7/4} M_X^{-3/4} \simeq 53.28 ~{\rm GeV}.
\end{equation}
Given that the LHC has not seen any evidence of supersymmetry up to now, 
we can set $M_R \geq 1$ TeV.  In that case, $M_X \geq 50$ TeV (the dashed 
line in Fig. \ref{fig:unify} is included to easily observe the change of slope 
at $M_X$ in the running of $\alpha_X$).  As a result,
interactions involving $\zeta_{1,2,3}$ may be ignored in our studies of 
dark matter.  We further assume that the $N_{1,2,3}$ singlets are also 
heavy, so they may also be ignored.  

For our scenario, we assume 
the masses $m_\chi, m_n, m_\eta$ of the three stable dark-matter particles 
$\tilde{\chi}^0_1, n, \tilde{\eta}^0_{R}$ to be arranged in ascending order. 
$\tilde{\eta}^0_{R}$ has $I_{3L}=0$, so it couples only to $Z'$.  Hence 
the annihilation of $\tilde{\eta}^0_{R} \bar{\tilde{\eta}}^0_{R}$ to $Z'$ 
to particles with masses less than $m_\eta$ will determine its relic abundance. 
Once $\tilde{\eta}^0_{R}$ freezes out, we need to consider the interactions of 
$n$.  Again $n$ has $I_{3L}=0$, so it couples to $Z'$, but 
there is also the interaction $\bar{e} n^c W_R^-$.  Hence the annihilation 
of $n \bar{n}$ occurs through $Z'$ to particles with mass less than $m_n$ 
as well as to $e^+e^-$ through $W_R^\pm$ exchange.  This will determine 
the relic abundance of $n$.  After $n$ freezes out, the remaining 
particles are presumably those of the MSSM, and the annihilation of 
$\tilde{\chi}^0_1 \tilde{\chi}^0_1$ will determine its relic abundance. 


This added flexibility should relax some of the most 
stringent constraints facing the MSSM today.


\subsection{Bound on $Z'$ from LHC data}

\begin{figure}[thb]
$$
\includegraphics[height=5.5cm]{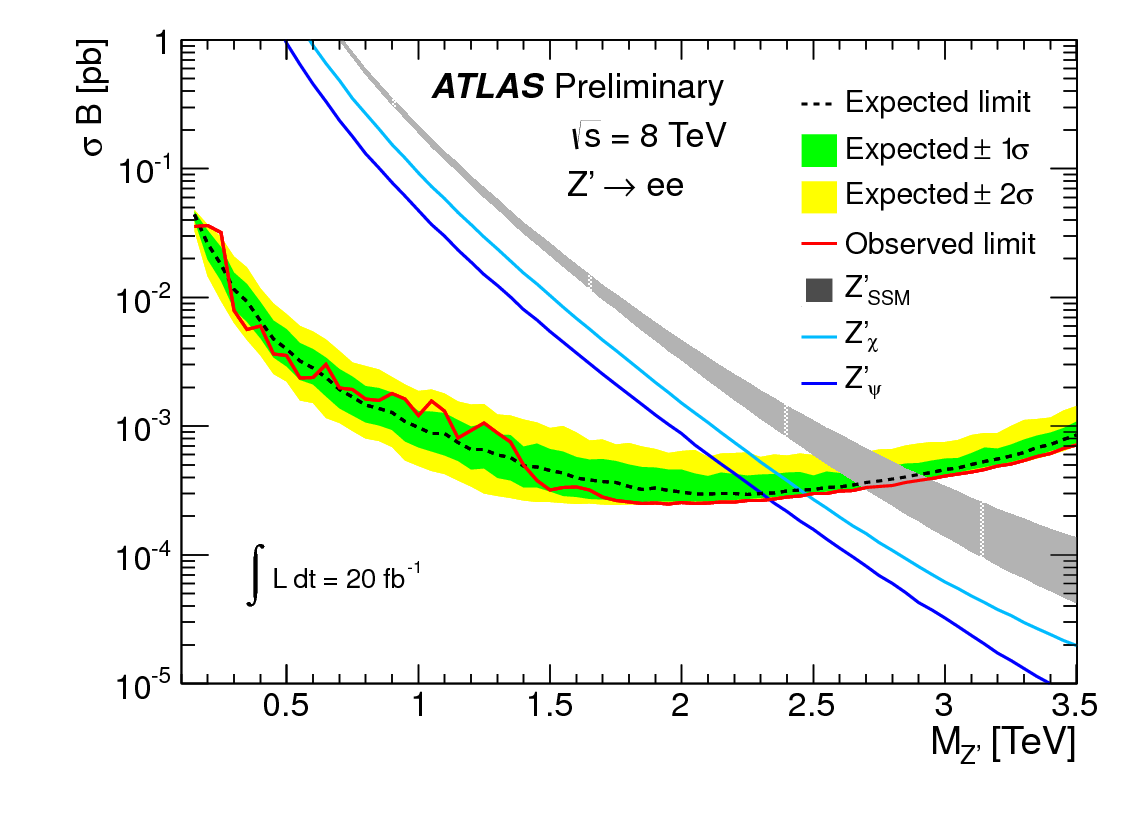}
\includegraphics[height=5.5cm]{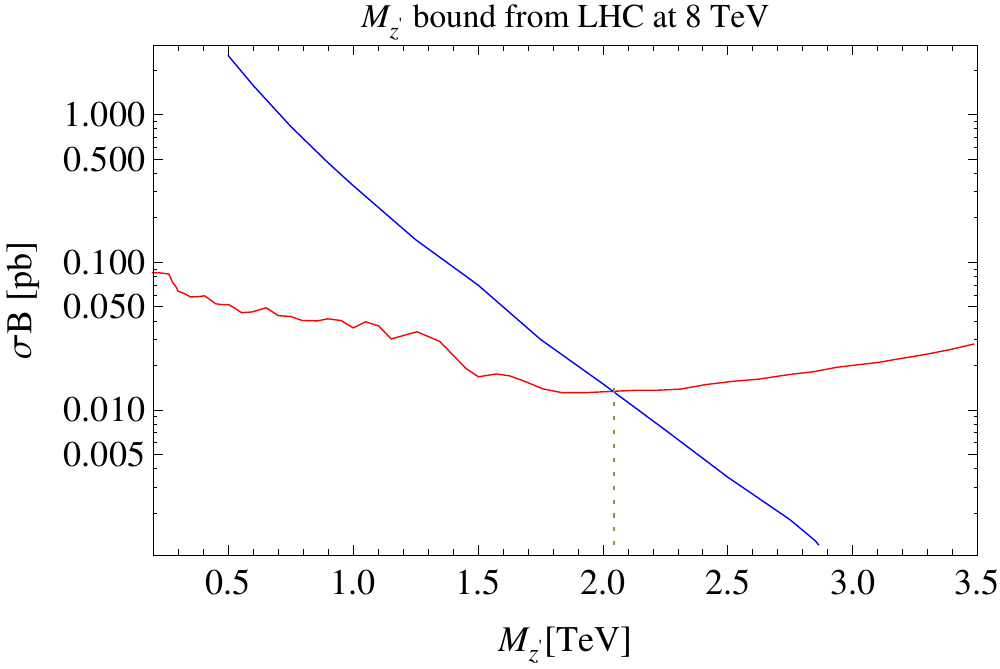}
$$
\caption{LHS: Bound on different $Z'$ masses at LHC from ATLAS with $E_{CM}=$ 8 TeV and integrated luminosity
of 20 $fb^{-1}$. RHS: The limit is exploited to determine the bound on the $Z'$ mass of this model. }
\label{zpbound}
\end{figure}

 $Z'$ couples to the current~\cite{klm09,klm10}
\begin{equation}
J_{Z'} = s_R^2 J_{3L} + c_L^2 J_{3R} - s_R^2 J_{em},
\end{equation}
with strength $g_{Z'} = e/s_R c_L \sqrt{c_L^2-s_R^2}$.
Given the unification requirements in ~\cite{m10}, we assume, $g_L=g_R$ which implies, $\sin\theta_R= \sin\theta_L \equiv \sin\theta_W$. 
We evaluate the bound on the mass of $Z'$ in our model from LHC data with $E_{CM}=$ 8 TeV and 
 integrated luminosity of 20 $fb^{-1}$. The result is shown in Fig \ref{zpbound}. On the left, we show the figure from 
 ATLAS \cite{ATLAS1}, where the bound was obtained for producing $Z'$ and subsequent decays to $e^{\pm}e^{\mp}$ 
 for some popular $Z'$ models. Right hand side shows our model cross-sections in blue and the bound 
 from LHC data in red, as seen in the LHS of the figure. 
 The cuts on the electron $p_T >40$ GeV and pseudorapidity $|\eta|< 2.47$ has been employed to obtain the signal in our model. 
 We use event generator {\tt CalcHEP} \cite{calchep} for calculating the cross-section and use {\tt CTEQ6L} parton distribution function \cite{CTEQ}. 
 From Fig. \ref{zpbound} we obtain the bound on the mass of $Z'$, $M_{Z'}=2.045 $ TeV $\simeq$ 2 TeV. 
 The bound on SSM, the phenomenological $Z'$ model with SM coupling, 
 has been cross-checked to be around 2.8 TeV, as shown on the left hand side.
 
Gauge Boson masses are calculated to be:
\begin{eqnarray}
M_{Z} &=& \frac{g_{L}v_{L}}{\sqrt{2(1-2\sin^{2}\theta_{W})}} \\ \nonumber
M_{Z'}&=& M_Z \sqrt{\sin^2 \theta_W + r^2 \cos^2 \theta_W} \\
M_{W_R}&=&M_Z \sqrt{\sin^2 \theta_W + r^2 (\cos^2 \theta_W-\sin^2 \theta_W)} \nonumber
\label{eqn:m-r}
\end{eqnarray}

\begin{figure}[thb]
$$
\includegraphics[height=6.5cm]{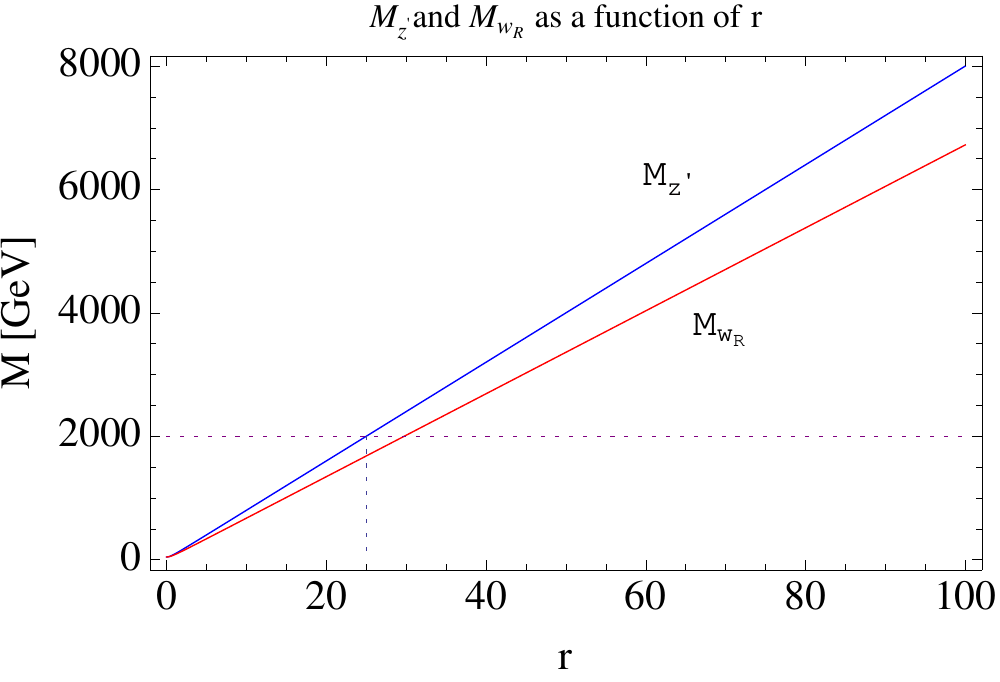}
$$
\caption{Linear dependence of $M_{Z'}$ (Blue) and $M_{W_R}$ (Red) on the ratio of 
Higgs vevs $r$ as defined in Eqn.\ref{eqn:r}. A horizontal dotted line indicates the bound from LHC on $Z'$ mass at 2 TeV.}
\label{m-r}
\end{figure}

Where $(v_L^2/u^2)=(1-2\sin^2 \theta_W)/ \sin^2\theta_W$ has been used to assume zero $Z-Z'$ mixing, and we have defined 
the ratio of Higgs vacuum expectation values as:
\begin{equation}
r=\frac{v_R}{v_L}
\label{eqn:r}
 \end{equation}
with $v_L^2= v_{L1}^2+v_{L2}^2, v_R^2= v_{R1}^2+v_{R2}^2$ and $ u^2= u_1^2+u_4^2$.

In Fig. \ref{m-r}, we show the linear dependence of the $Z'$ and $W_R$ mass on the ratio of Higgs vacuum expectation values
$r$ following Eqn. \ref{eqn:m-r}. We note that mass of $Z'$ is bigger than $W_R$ for $M_Z' \ge$ 30 GeV. 
The bound on $M_Z' \ge$ 2 TeV from LHC eventually put a bound on $r \ge$ 25 as shown. In the following analysis, 
we use $r$ as a plotting variable instead of $M_Z'$ or $M_{W_R}$.

\subsection{Relic Abundance of $n$ and $\tilde{\eta}^0_R$}

\begin{figure}[thb]
\begin{center}
\begin{tikzpicture}[line width=1.5 pt, scale=1.3]
        \draw[fermionbar] (-140:1)--(0,0);
        \draw[fermion] (140:1)--(0,0);
        \draw[vector] (0:1)--(0,0);
        \node at (-140:1.2) {$\tilde{\eta}_R^0$};
        \node at (140:1.2) {$\tilde{\eta}_R^0$};
        \node at (.5,.3) {$Z'$};       
\begin{scope}[shift={(1,0)}]
        \draw[fermion] (-40:1)--(0,0);
        \draw[fermionbar] (40:1)--(0,0);
        \node at (-40:1.2) {$f$};
        \node at (40:1.2) {$f$};       
\end{scope}
\end{tikzpicture}
\end{center}
\caption{Feynman Diagram for $\tilde{\eta}_R^0$ annihilation to SM.}
\label{ann1}
\end{figure}
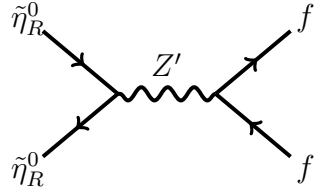

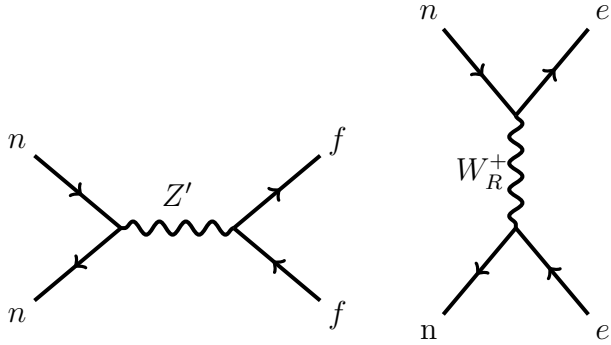
\begin{figure}[thb]
\begin{center}
\begin{tikzpicture}[line width=1.5 pt, scale=1.5]
        \draw[fermionbar] (-140:1)--(0,0);
        \draw[fermion] (140:1)--(0,0);
        \draw[vector] (0:1)--(0,0);
        \node at (-140:1.2) {$n$};
        \node at (140:1.2) {$n$};
        \node at (.5,.3) {$Z'$};       
\begin{scope}[shift={(1,0)}]
        \draw[fermion] (-40:1)--(0,0);
        \draw[fermionbar] (40:1)--(0,0);
        \node at (-40:1.2) {$f$};
        \node at (40:1.2) {$f$};       
\end{scope}
\begin{scope}[shift={(3.5,0)}]
           \begin{scope}[rotate=90]
                        \draw[fermion] (-140:1)--(0,0);
                        \draw[fermionbar] (140:1)--(0,0);
                        \draw[vector] (0:1)--(0,0);
                        \node at (-140:1.2) {$e$};
                        \node at (140:1.2) {n};
                        \node at (.5,.3) {$W_R^+$};    
                \begin{scope}[shift={(1,0)}]
                        \draw[fermionbar] (-40:1)--(0,0);
                        \draw[fermion] (40:1)--(0,0);
                        \node at (-40:1.2) {$e$};
                        \node at (40:1.2) {$n$};       
                \end{scope}
        \end{scope}
        \end{scope}
\end{tikzpicture}
\end{center}
\caption{Feynman Diagram for $n$ annihilation to SM.}
\label{ann2}
\end{figure}

The annihilation cross-sections for DM $\tilde{\eta}_R^0$ to SM particles goes through s-channel diagram exchanging $Z'$, 
while $n$ has an additional piece through a t-channel diagram to $e_R^{\pm}$ through $W_R^{\pm}$ exchange.
The Feynman diagrams are shown in fig. \ref{ann1} and  \ref{ann2}. 

The expressions for thermally averaged cross-section ($\langle \sigma v \rangle$)  for these two DM components annihilating to SM are indicated in 
Eqn. \ref{eqn:sigmav-eta} and Eqn. \ref{eqn:sigmav-n}. 
\begin{eqnarray}
\langle \sigma v \rangle_{\eta}  &\simeq & 
\frac{g_{L}^4}{64 \pi} \frac{m_{\tilde{\eta}_R^0}^2}{(4m_{\tilde{\eta}_R^0}^2-M_{Z'}^2)^2} \frac{(10 - 29 c_W^2 +22 c_W^4)}{(2c_W^2-1)^2}
\label{eqn:sigmav-eta}
\end{eqnarray} 

\begin{eqnarray}
 \nonumber & &\langle \sigma v \rangle_{n} \simeq \frac{g_R^4 m_n^2}{64 \pi}  \bigg[ \frac{10 s_W^4 - 9 s_W^2 c_W^2 +3c_W^4 }{(4m_n^2- M_{Z'}^2)^2(c_W^2-s_W^2)^2} \\ 
 & & + \frac{3}{(m_n^2 + M_{W_R}^2)^2} + \frac{3 (c_W^2 - 2 s_W^2)}{(4m_n^2- M_{Z'}^2)(m_n^2 + M_{W_R}^2)(c_W^2-s_W^2)} \bigg]
 \label{eqn:sigmav-n}
\end{eqnarray}

With the unification condition, $g_R^2 =g_L^2 \simeq 0.427 $ and  $\sin^2 \theta_W=0.23$, numerically, we obtain:
 \begin{eqnarray}
{\langle \sigma v \rangle}_\eta \simeq \frac{0.00222 {m_{\tilde{\eta^0_R}}}^2}{4 {m_{\tilde{\eta^0_R}}}^2-M_Z'^2 }
\label{eqn:sigma-v-eta2}
\end{eqnarray} 

\begin{eqnarray}
{\langle \sigma v \rangle}_n \simeq \frac{0.00222 {m_n}^2}{4 {m_n}^2-M_Z'^2 }+\frac{0.00272 {m_n}^2}{{m_n}^2+{M_{W_R}}^2 }+
\frac{0.00156 {m_n}^2}{(4 {m_n}^2-M_Z'^2)({m_n}^2+{M_{W_R}}^2 ) }
\label{eqn:sigma-v-n2}
\end{eqnarray} 

If we assume the decoupling of ${\tilde{\eta}_R^0}$, $n$ and $\tilde{\chi}_1^0$ from the hot soup of SM particles 
are independent of interactions with each other, relic density for each DM component can be approximated as
\begin{eqnarray}
\Omega_{i} h^2 \simeq \frac {0.1 pb} {\langle \sigma v \rangle}_i
\label{eqn:omega}
\end{eqnarray}
The total abundance will be a sum of the three DM components, i.e.
\begin{eqnarray}
{\Omega_{DM_{tot}} h^2} = {\Omega_{\eta} h^2} +{\Omega_{n} h^2} +{\Omega_{\chi_1^0}  h^2}
\label{eqn:omegatot}
\end{eqnarray}

\begin{figure}[thb]
$$
\includegraphics[height=5.5cm]{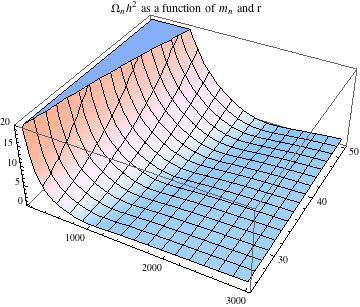}
\includegraphics[height=5.5cm]{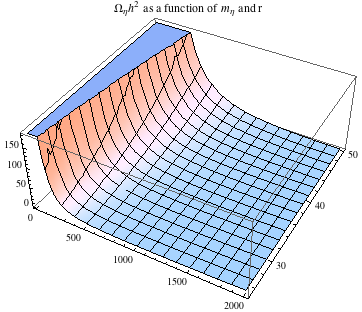}
$$
\caption{A 3-dim plot showing $\Omega h^2$ (z-axis) dependence on mass  (x-axis) and $r$ (y-axis). LHS: $n$ and RHS: $\tilde{\eta}_R^0$}
\label{Omega-m-r}
\end{figure}

\begin{figure}[thb]
$$
\includegraphics[height=5.5cm]{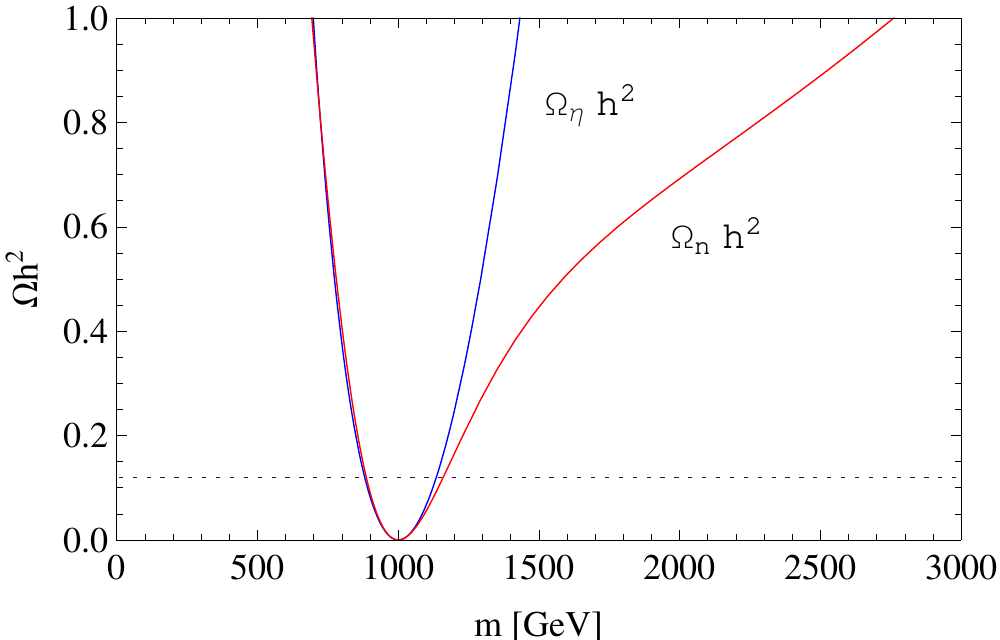}
$$
\caption{$\Omega h^2$ dependence for DM $n$ and ${\tilde{\eta}_R^0}$ on mass for $r=$25.}
\label{Omega-m1}
\end{figure}

With this assumption, we evaluate relic abundance for each of the DM component and look for the parameter space
where they add up to the constraint from WMAP \cite{WMAP} \footnote{ PLANCK \cite{PLANCK} data essentially indicates 
a very similar range, though more stringent, almost indistinguishable from WMAP in present context. }. 
\begin{eqnarray}
0.094<{\Omega_{DM_{tot}} h^2} < 0.130
\label{eqn:omegatot2}
\end{eqnarray}

In Fig. \ref{Omega-m-r}, we show a 3-dimensional plot with $\Omega h^2$ along z-axis, DM mass 
$m$  along x-axis and the ratio of Higgs vevs $r$ along y-axis for the DM component $n$ on LHS and ${\tilde{\eta}_R^0}$ on RHS. We use 
Eqn \ref{eqn:sigma-v-eta2}, Eqn \ref{eqn:sigma-v-n2} and Eqn \ref{eqn:omega} to draw them. Both of the DMs show similar behaviour. Now, a cut along the $r$-axis 
at 25, shows the dependence of $\Omega h^2$  on DM mass $m$ which is shown in Fig \ref{Omega-m1}. The difference 
in $n$ and ${\tilde{\eta}_R^0}$ annihilation is clear from here.   

\begin{figure}[htbp]
\begin{center}
\includegraphics[height=7cm]{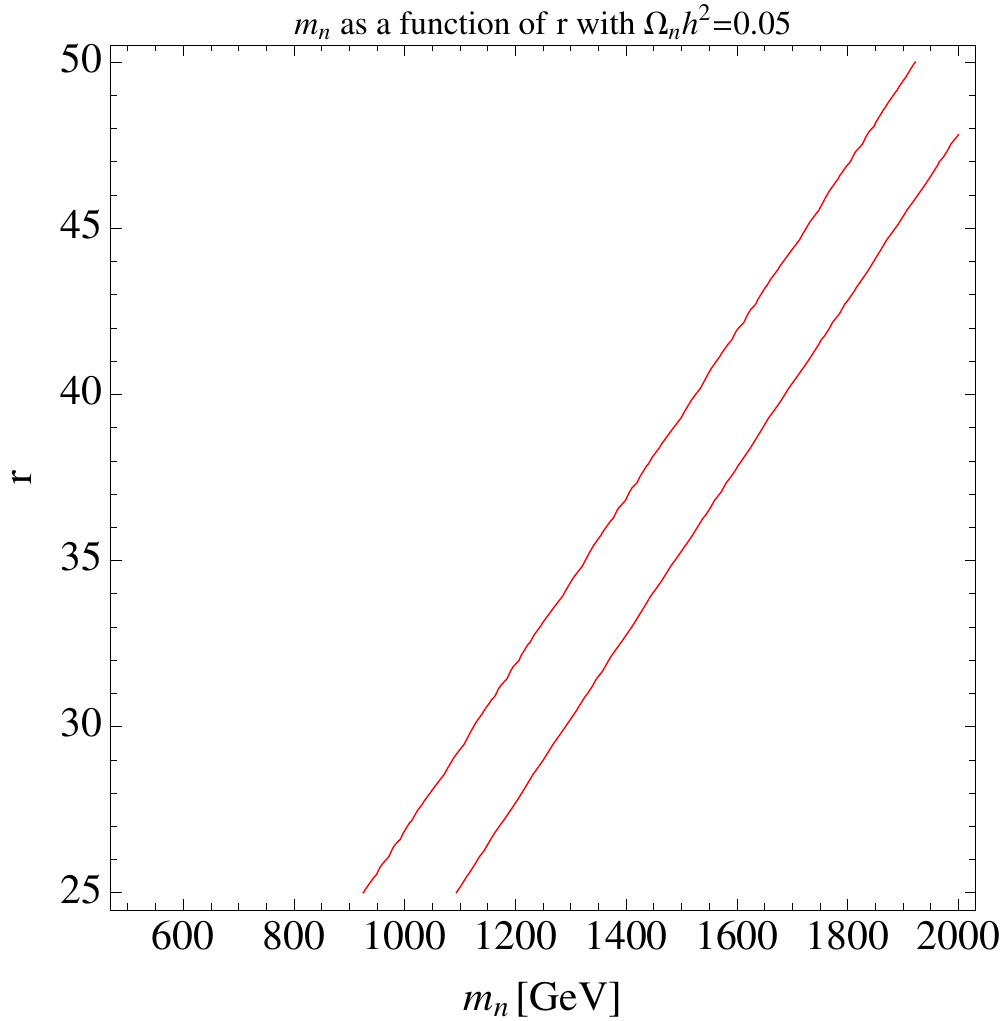}
\hskip 20 pt
\includegraphics[height=7cm]{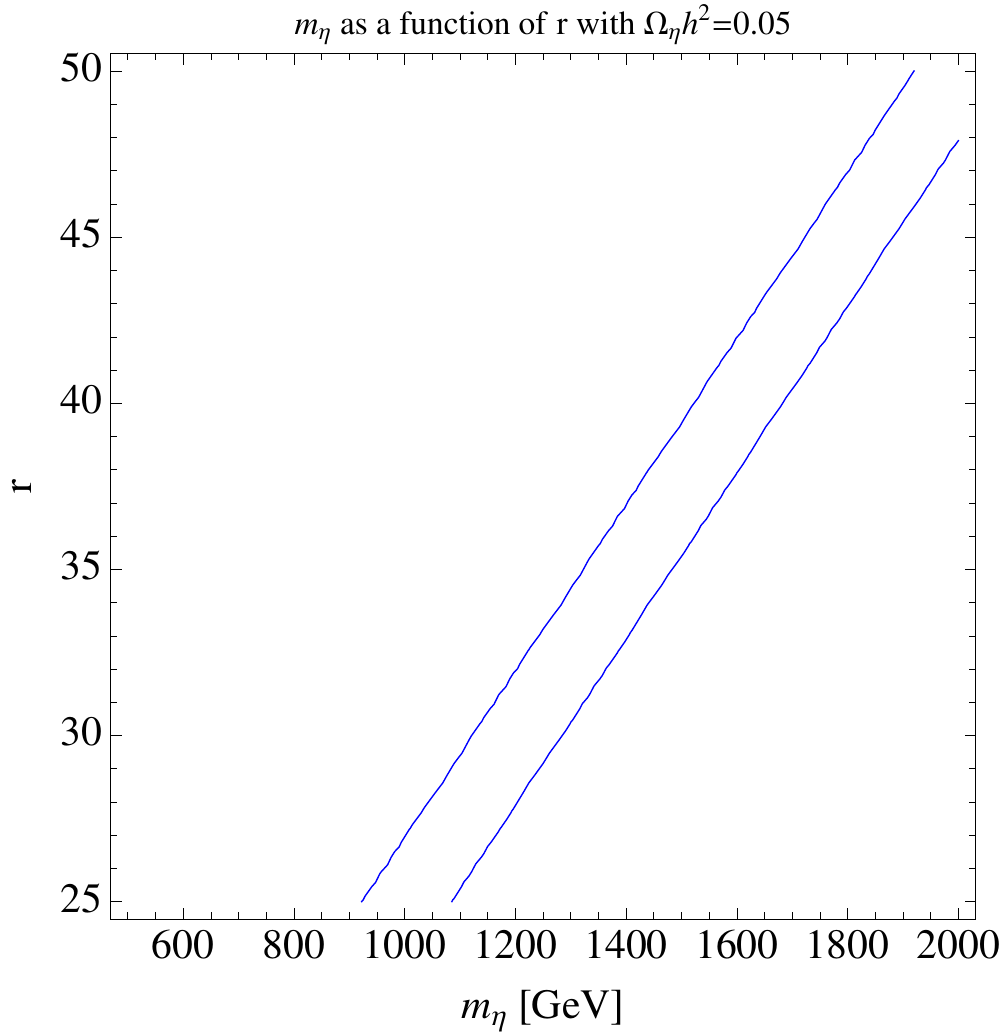}
\vskip 10 pt
\includegraphics[height=7cm]{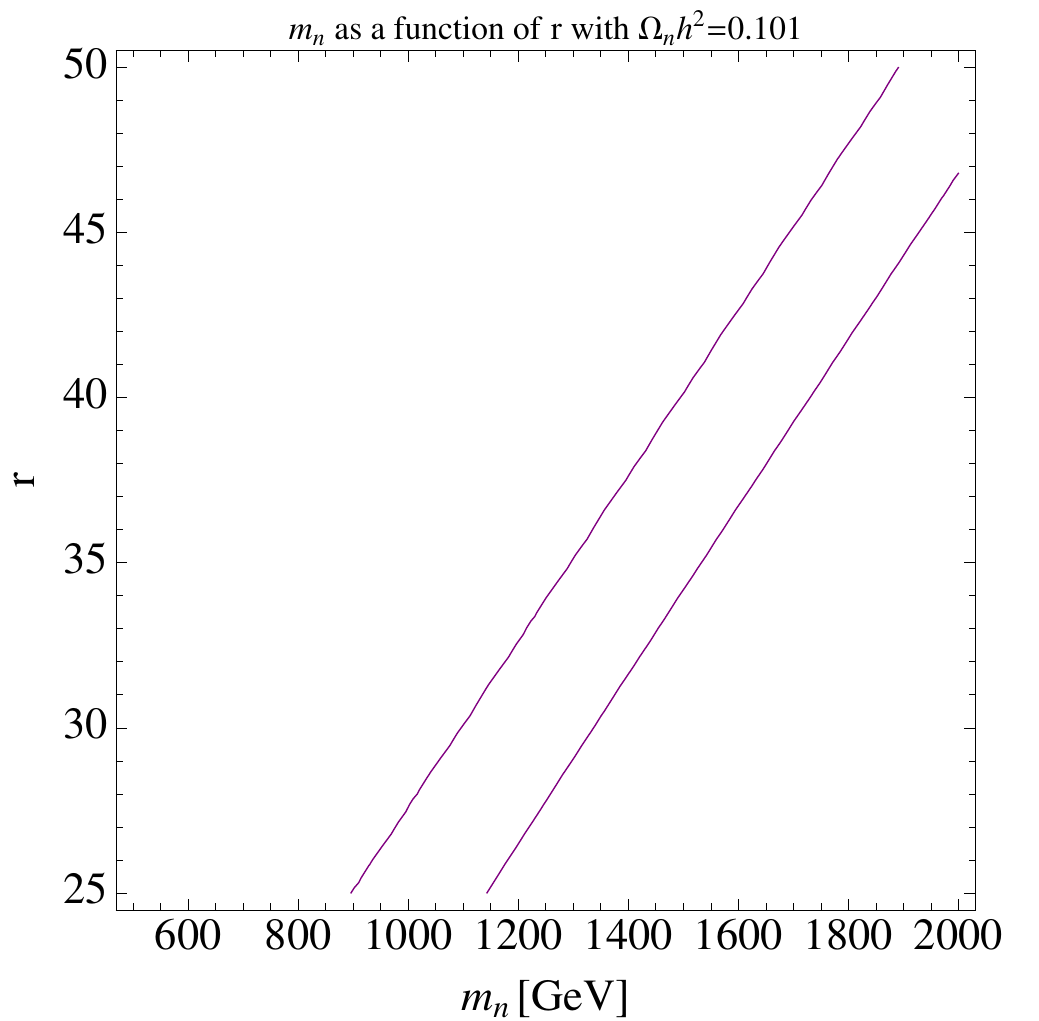}
\caption{Top: $m-r$ dependence when each of the DM component $n$ (left) and ${\tilde{\eta}_R^0}$ (right) DM contributing 
equally to the relic density with $\Omega_i h^2=$0.05. Bottom: When one component dominates, i.e. $\Omega_i h^2=$0.1. Masses are in GeVs.}
\end{center}
\label{fig:m-r-2}
\end{figure} 

In the three component DM framework, we study a scenario where the two components $n$ and $\tilde{\eta}_R^0$ dominate in relic abundance
leaving a very tiny space for neutralino $\tilde{\chi}_1^0$. We will discuss neutralino DM shortly. For example, we focus on a region of parameter space, where,
\begin{eqnarray}
{\Omega_{\eta} h^2} + {\Omega_{n} h^2} = 0.1
\label{eqn:2-comp}
\end{eqnarray}

In such a case, if we assume in addition that each of the components contribute equally, then we end up getting Fig. 9. 
This indicates that we obtain two possible masses for a given value of $r$ and $\Omega h^2$ and the difference in $n$ and $\tilde{\eta}_R^0$
annihilation doesn't matter in the range of $r$ and $\Omega h^2$ we are interested. This is shown in the top panel of Fig. 9, 
for $n$ (left) and $\tilde{\eta}_R^0$ (right). They look exactly the same, where DM mass is plotted with $r$. In the bottom panel, 
we show the case when one of the components contribute fully to relic abundance with $ {\Omega_{i} h^2} = 0.1$. 

\begin{figure}[thb]
$$
\includegraphics[height=7cm]{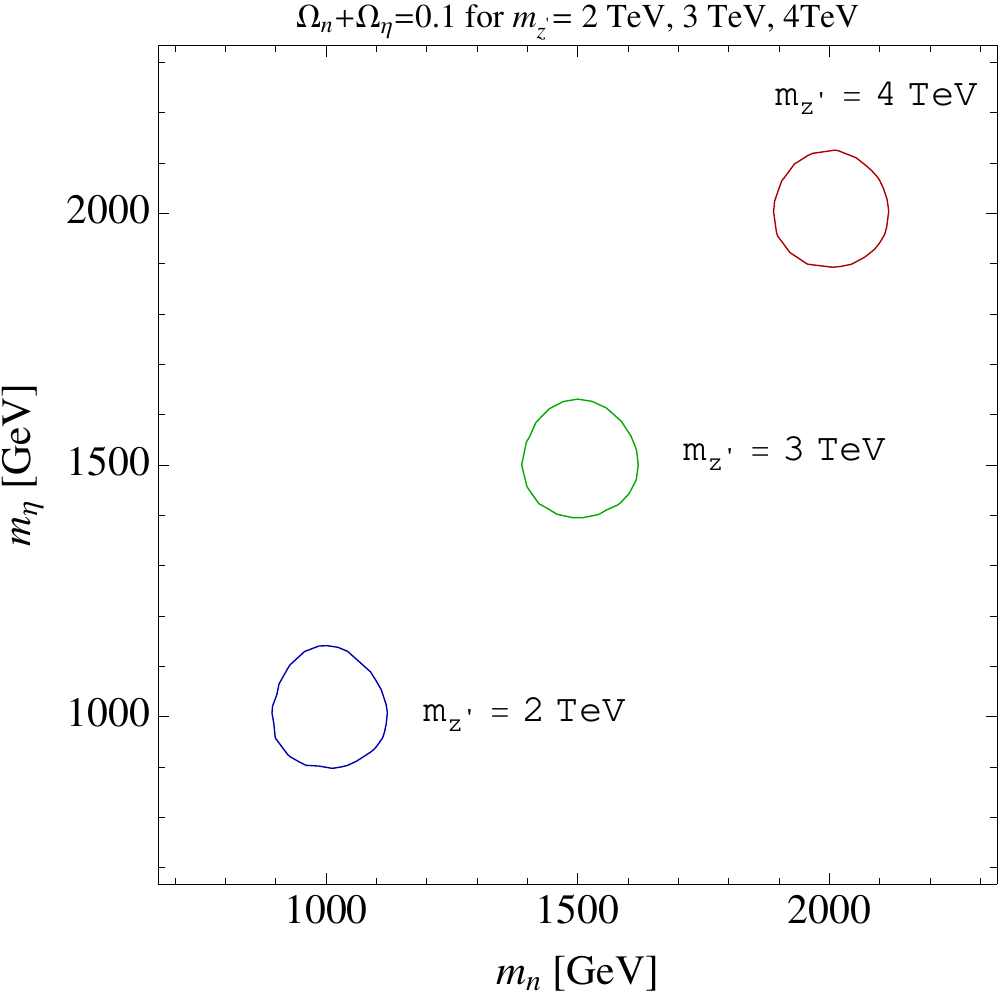}
\hskip 20 pt
\includegraphics[height=7cm]{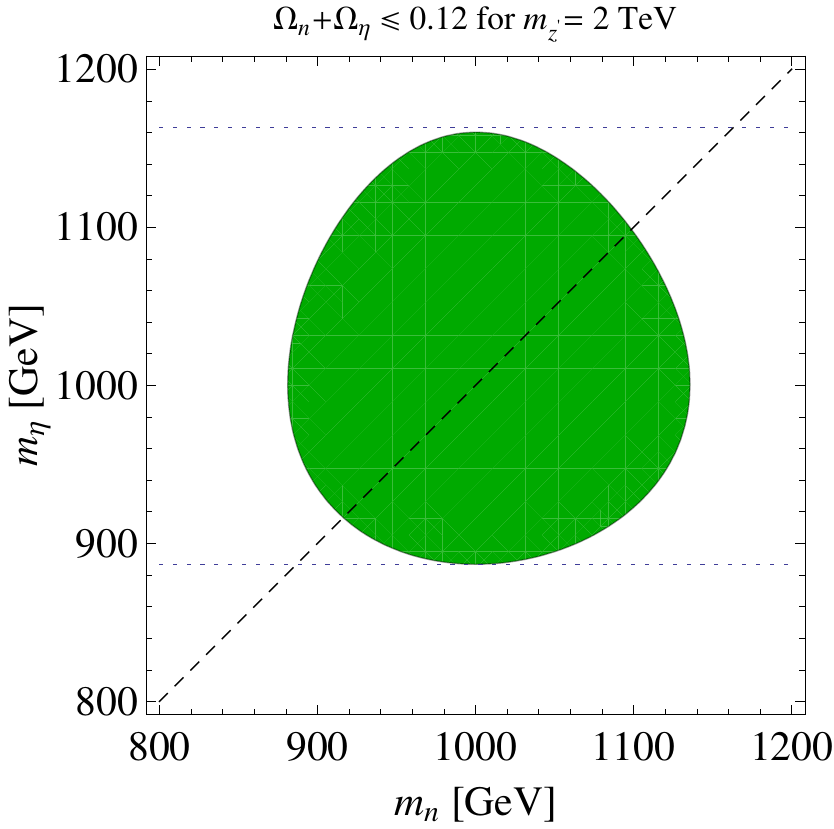}
$$
\caption{LHS: A plot showing $m_n$-$m_\eta$ [GeV] contours for $\Omega_n h^2+ \Omega_\eta h^2$=0.1 for $M_Z' =$ 2, 3, 4 TeV. 
RHS: Region of the $m_n$-$m_\eta$ [GeV] parameter space when $\Omega_n h^2+ \Omega_\eta h^2 \le$0.12 with $M_Z'$= 2 TeV.}
\label{contour}
\end{figure}

Eqn. \ref{eqn:2-comp} is appropriately depicted in Fig. \ref{contour} for different $Z'$ masses. They represent as three 
circles (The circular shape is understandable from looking at Fig. \ref{Omega-m1}) in $m_n$ and $m_{\tilde{\eta}_R^0}$ [GeV] plane for $M_Z' = $ 2, 3 and 4 TeV around $m_n=m_{\tilde{\eta}_R^0}=M_Z'/2$.  
The reason is simple to understand; the resonance region essentially contributes for relic abundance. We 
highlight the case for $M_Z' = $ 2 TeV in the RHS of Fig. \ref{contour}. The whole region in green becomes allowed 
when we have the condition ${\Omega_{\eta} h^2} + {\Omega_{n} h^2} \le$ 0.12 (i.e. the contour shrinks 
for smaller abundance). We also note that, if we adhere to the assumption 
made initially that $m_{\eta} \ge m_n$, then only half of the circle above the diagonal line is allowed for 
relic abundance restricting the allowed mass range for $n$ between 866-1100 GeV and for $\tilde{\eta}_R^0$ between 915-1163 GeV. 
Given that the plot is close to a perfect circle, $\Omega_n h^2 \ge \Omega_\eta h^2$ in this limit. Hence,
if $n$ and $\tilde{\eta}_R^0$ together contributes to 90$\%$ of the total dark matter relic density, $\tilde{\eta}_R^0$ can contribute 
in 1- 45$\%$ and $n$ can contribute in 45-90$\%$. 

\begin{figure}
\begin{center}
\begin{tikzpicture}[line width=1.5 pt, scale=1.5]
        \begin{scope}[rotate=90]
                        \draw[fermion] (-140:1)--(0,0);
                        \draw[fermionbar] (140:1)--(0,0);
                        \draw[vector] (0:1)--(0,0);
                        \node at (-140:1.2) {q};
                        \node at (140:1.2) {q};
                        \node at (.5,.3) {$Z'$};       
                \begin{scope}[shift={(1,0)}]
                        \draw[fermionbar] (-40:1)--(0,0);
                        \draw[fermion] (40:1)--(0,0);
                        \node at (-53:1.2) {$n,\tilde{\eta^0_R}$};
                        \node at (53:1.2) {$n,\tilde{\eta^0_R}$};    
                \end{scope}
                 \end{scope}
\end{tikzpicture}
\caption{Diagram for scattering with quarks for direct detection.}
\label{direct}
\end{center}
\end{figure}
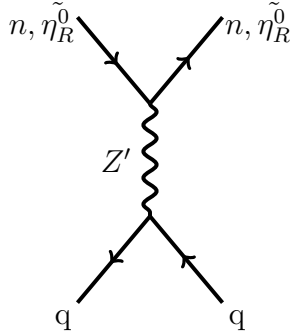

\subsection{Direct Detection of $n$ and $\tilde{\eta}^0_R$}

Direct detection of $n$ and $\tilde{\eta}_R^0$ takes place through t-channel $Z'$ interaction with quarks. 
The Feynman graph is shown in fig. \ref{direct}. Due to only this contribution, the spin-independent (SI) cross-section is very small.

We use {\tt MicrOMEGAs} \cite{MicrOmega} to calculate
the effective SI nucleon scattering cross-section.  
The parton-level interaction is converted to the nucleon level by using 
effective nucleon ${f_q}^N$ $(N = p,n)$ couplings defined as ~\cite{MicrOmega}
\begin{eqnarray}
\langle N|m_q\bar{\psi_q}\psi_q|N \rangle={f_q}^N M_N \,,
\end{eqnarray}
where $M_N$ is the nucleon mass and we use the default form factors in ~\cite{MicrOmega} as
$f_u^p= 0.033,\; f_d^p= 0.023,\; f_s^p= 0.26$, for the proton;  $f_u^n= 0.042,\; f_d^n= 0.018,\;f_s^n= 0.26$
for the neutron; while for the heavy quarks the $f_q^N$ are generated by gluon exchange with the nucleon and are given by
 \begin{eqnarray}
f_Q^N=\frac{2}{27} \left (1-\sum_{q=u,d,s} f_q^N  \right ) \, \quad Q=c,t,b.
\label{eq:fQ}
\end{eqnarray}

\begin{figure}[thb]
$$
\includegraphics[height=7cm]{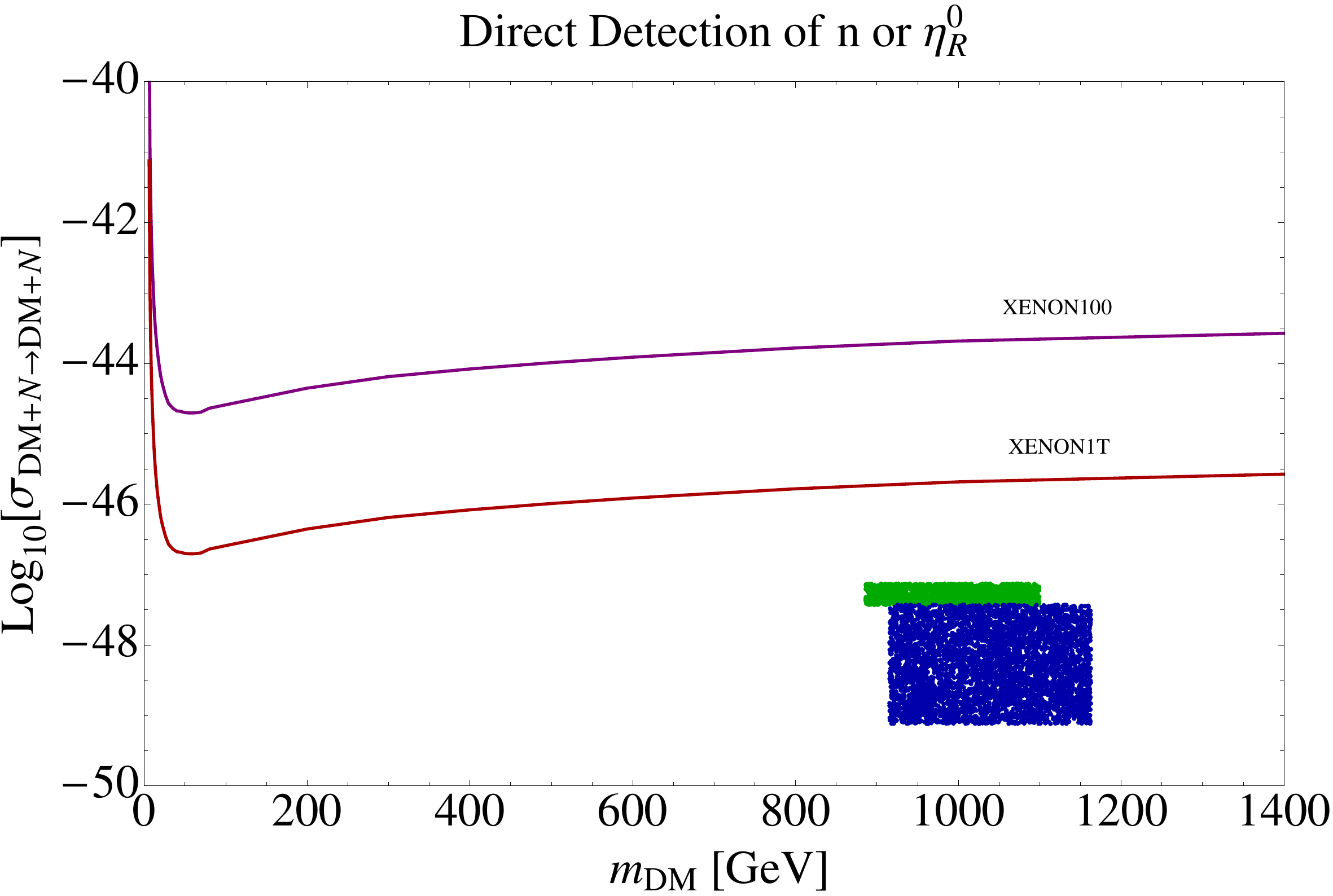}
$$
\caption{Direct detection constraint for DM $n$ and $\tilde{\eta}_R^0$. Spin-independent effective nucleon 
cross-section [cm$^2$] in log scale is plotted in y-axis as a function of DM mass [GeV] along x-axis. 
The upper thick curve in purple shows the limit from XENON100 and the lower one in red is for XENON1T. 
Points in the blue box represents $\tilde{\eta}_R^0$ contributing 1-45 $\%$ (bottom to top) and 
those in green correspond to $n$ contributing 45-90 $\%$ (bottom to top) of the total dark matter density in 
WMAP allowed mass range.}
\label{DDn}
\end{figure}

The results are shown in Fig. \ref{DDn}. The bounds from XENON100 (above) and 
XENON1T (below) are shown in two continuous lines in purple and red respectively. Any points above the XENON100 
lines will be discarded by the direct search experiments. In Fig. \ref{DDn}, points in blue shows the results of  
SI direct detection cross-section  for $\tilde{\eta}_R^0$ with $M_Z'$= 2 TeV  
 and those in green represent $n$ within the allowed mass range to obtain correct relic density; $m_n$ 
 between 866-1100 GeV and for $m_{\eta}$ between 915-1163 GeV. 
 Although $n$ and $\tilde{\eta}_R^0$ have same quark interaction 
as in Fig. \ref{direct} and have same direct detection cross-section, given the mass hierarchy 
$m_{\eta} \ge m_n$, $n$ contributes more than $\tilde{\eta}_R^0$ to the dark matter density. 
Due to multi-component nature of the dark matter, the effective direct detection cross-section for each DM component 
is obtained by multiplying the fraction of their number density $ \frac{n_{DM}}{n_{tot}}$ 
with the actual nucleon cross-section $\sigma_N$ (assuming that all of the DMs are accessible to the detector). 

\begin{equation}
{\sigma_{N}}_{eff}= \frac{n_{DM}}{n_{tot}} \sigma_N \simeq \frac{\Omega_{DM} h^2}{\Omega_{tot}h^2} \sigma_N
\end{equation}

The thickness of the direct detection cross-section essentially comes from the fraction
$ \frac{n_{DM}}{n_{tot}}$, which has been varied between 1-45 $\%$ for $\tilde{\eta}_R^0$ 
(in blue) and 45-90$\%$ for $n$ (in green). Hence, points at the bottom of the blue box constitute 
only 1$\%$ while those at the top in green constitute 90$\%$  of the total DM. The unequal thickness
in blue and green box is due to the logarithmic scale of the effective cross-section.
 
The direct detection cross-section also doesn't depend on DM mass, while it depends 
on the $Z'$ mass very much. With higher $M_Z'$ they go down even below to make it harder for direct search. 
Possibility of early discovery of these DMs in near-future experiments seems to be small, 
although they are surely allowed by the exclusion limits set by XENON.

\subsection{Relic Abundance and Direct detection of Wino type of Neutralino $\tilde{\chi}_1^0$}

Let us now discuss the lightest neutralino ($\tilde{\chi}_1^0$) as the third DM candidates
in this three-component DM set up. The neutralino sector in this extended LR SUSY model is 
non-trivial and constitutes of three gauginos ($M_B$, $M_L$, $M_R$) and thirteen Higgsinos. Seven out of them, which are 
superpartners of the scalar fields that do not have a vev, do not mix with the gauginos or the rest of the Higgsinos. This yields 
to a nine dimensional neutralino mass matrix.   
\begin{figure}[thb]
$$
\includegraphics[height=7cm]{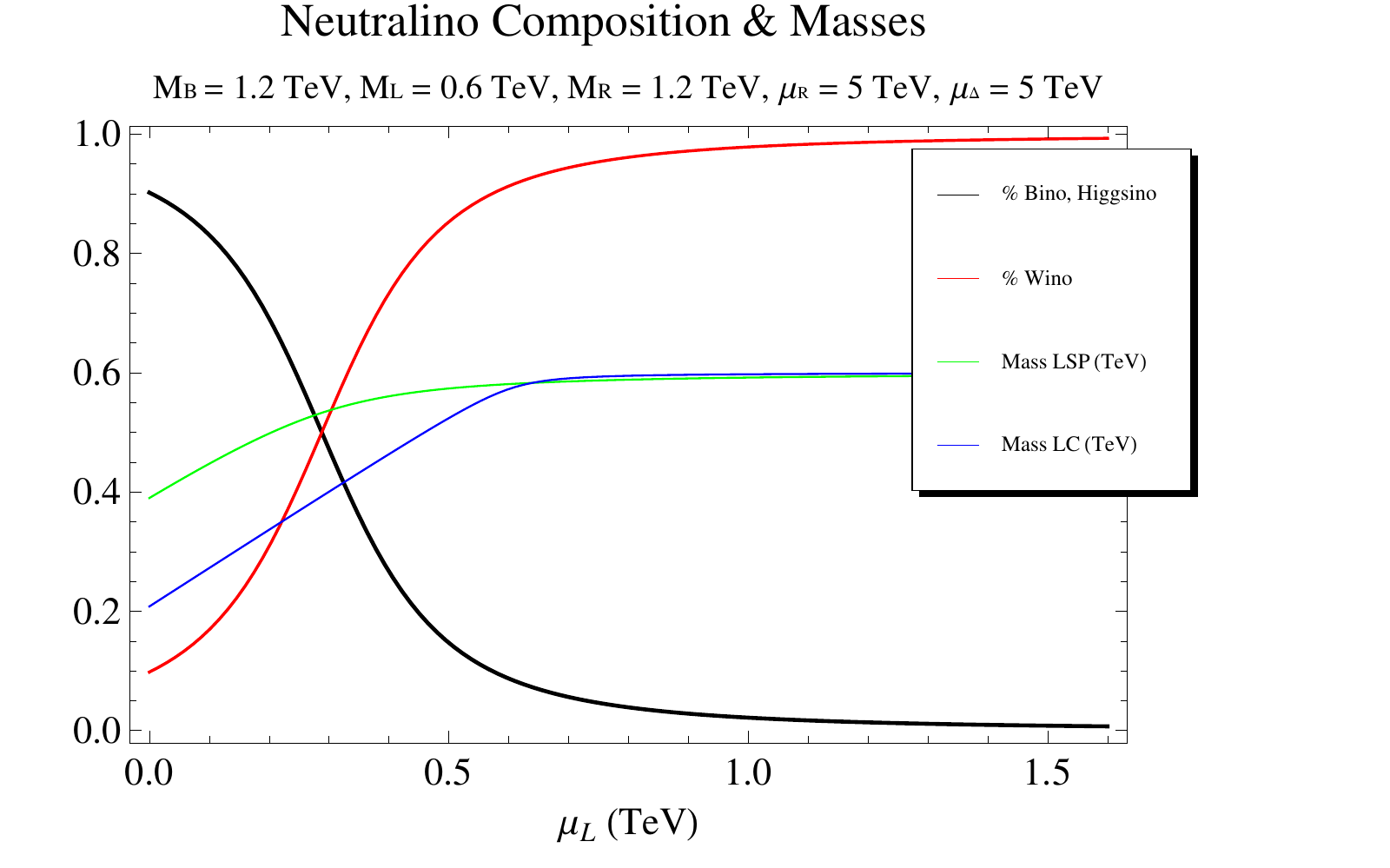}
$$
\caption{Plot showing a limit when the lightest neutralino becomes predominantly a Wino and the first chargino becomes degenerate with LSP. }
\label{Wino-DM}
\end{figure}

For simplicity, we take a limit where the neutralino DM is predominantly a wino. In this limit, the neutralino of this model can easily 
mimic minimal supersymmetric Standard Model (MSSM) neutralino, 
with $M_{B}=M_{1},\mu_{L}=\mu,\beta_{L}=\beta,$and $1.43M_{L}=M_{2}$. This is explicitly shown in the Appendix.
In Fig. \ref{Wino-DM}, we show as an example, that when $\mu_L$ (x-axis) is larger than $M_L$ (which we set at 0.6 TeV), 
fraction of bino and Higgsino components in lightest neutralino, in black thick line goes to almost zero; 
giving rise to a wino DM with the red line reaching 1. We also show that the lightest chargino 
(in blue, called LC) becomes degenerate with the lightest neutralino (in green) 
and both have mass around 600 GeV in this particular point in parameter space. 
This degeneracy is a very well known feature of wino dominated neutralino in MSSM. Note that in order to achieve
 this limit in this model, we kept $M_R\simeq M_B$ and other non-MSSM parts heavy, $ \mu_R, \mu_{\Delta} = $ 5 TeV.

\begin{figure}[thb]
$$
\includegraphics[height=6.5cm]{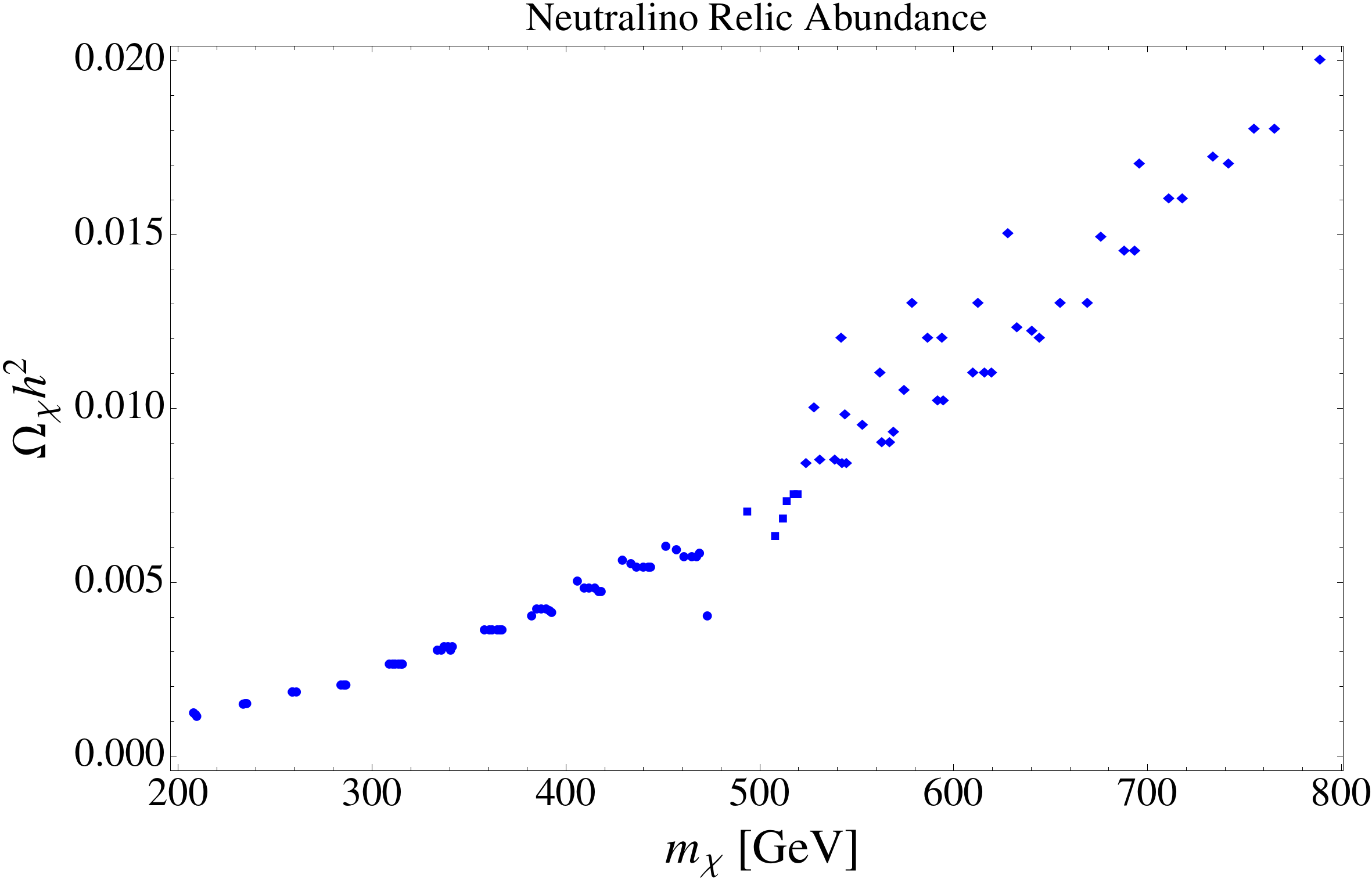}
$$
\caption{Relic Abundance of lightest neutralino as a function of mass when it is predominantly a wino. The scanned parameter space here ranges $M_1$: (800-1200) GeV, 
$M_2$: (200-775) GeV,  $\mu$: (600-1000) GeV with $\mu, M_1 > M_2$.}
\label{Omega-chi}
\end{figure}

\begin{figure}
\begin{center}
\begin{tikzpicture}[line width=1.5 pt, scale=1.5]
 \begin{scope}[rotate=90]
                        \draw[fermion] (-140:1)--(0,0);
                        \draw[fermionbar] (140:1)--(0,0);
                        \draw[vector] (0:1)--(0,0);
                        \node at (-140:1.2) {q};
                        \node at (140:1.2) {q};
                        \node at (.5,.3) {$Z$};       
                \begin{scope}[shift={(1,0)}]
                        \draw[fermionbar] (-40:1)--(0,0);
                        \draw[fermion] (40:1)--(0,0);
                        \node at (-53:1.2) {$\tilde{\chi}_1^0$};
                        \node at (53:1.2) {$\tilde{\chi}_1^0$};    
                \end{scope}
                 \end{scope}
                  \begin{scope}[shift={(3,0)}]
        \begin{scope}[rotate=90]
                        \draw[fermion] (-140:1)--(0,0);
                        \draw[fermionbar] (140:1)--(0,0);
                        \draw[dashed] (0:1)--(0,0);
                        \node at (-140:1.2) {q};
                        \node at (140:1.2) {q};
                        \node at (.5,.3) {$h$};       
                \begin{scope}[shift={(1,0)}]
                        \draw[fermionbar] (-40:1)--(0,0);
                        \draw[fermion] (40:1)--(0,0);
                        \node at (-53:1.2) {$\tilde{\chi}_1^0$};
                        \node at (53:1.2) {$\tilde{\chi}_1^0$};  
                        \end{scope}  
                \end{scope}
                 \end{scope}
                  \begin{scope}[shift={(6,0)}]
                 \begin{scope}[rotate=90]
                        \draw[fermion] (-140:1)--(0,0);
                        \draw[fermionbar] (140:1)--(0,0);
                        \draw[dashed] (0:1)--(0,0);
                        \node at (-140:1.2) {$\tilde{\chi}_1^0$};
                        \node at (140:1.2) {q};
                        \node at (.5,.3) {$\tilde{q}$};       
                \begin{scope}[shift={(1,0)}]
                        \draw[fermionbar] (-40:1)--(0,0);
                        \draw[fermion] (40:1)--(0,0);
                        \node at (-53:1.2) {q};
                        \node at (53:1.2) {$\tilde{\chi}_1^0$};   
                          \end{scope}
                \end{scope}
                 \end{scope}
\end{tikzpicture}
\caption{Diagram for lightest neutralino $\tilde{\chi}_1^0$ scattering with quarks for direct detection.}
\label{directChi}
\end{center}
\end{figure}
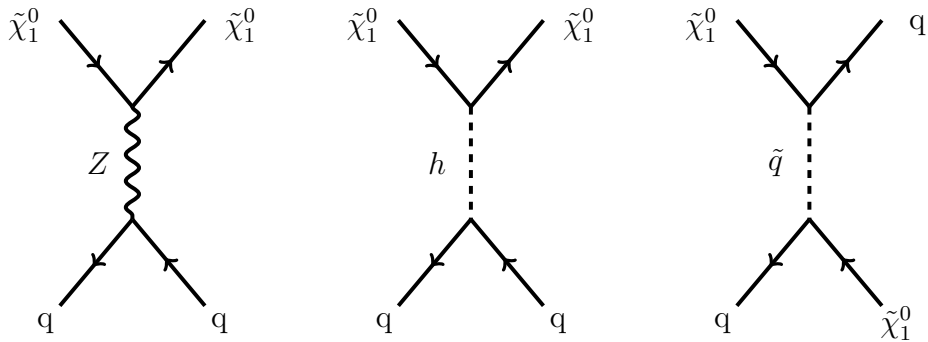

It is also known that when lightest chargino is degenerate with neutralino DM, co-annihilation occurs \cite{Griest:1990kh}, making $\tilde{\chi}_1^0$ 
annihilation cross-section much larger to yield very small abundance.  This has been crafted in different ways 
\cite{Nelson,Lisa,Baer,Debottam,Moroi} to make wino a viable DM candidate by having moduli decay 
in anomaly mediated SUSY breaking \cite{Lisa} or by non-thermal productions \cite{Moroi} etc. Wino DM has been studied also 
to justify PAMELA data \cite{Pamela}. However, the under-abundance works perfectly fine for us with
the other two components to make up. Of course, other regions of neutralino DM parameter space where it is an admixture of 
Higgsino-wino-bino that yields under-abundance is also allowed for the model. We show a sample scan of wino dominated neutralino 
for relic density and direct detection. The MSSM parameter space scanned here: $M_1$ between 800-1200 GeV, $M_2 \simeq 1.43 M_L$, between 200-775 GeV, 
and the Higgsino parameter $\mu$ between 600-1000 GeV (with $\mu, M_1 > M_2$). In Fig \ref{Omega-chi}, we show that the neutralino-DM 
under abundance for $\Omega_{\chi_1^0}h^2$ is not larger than 0.02 if we keep $m_{\tilde{\chi}_1^0} \le$ 800 GeV 
(This is following the assumption that neutralino is the lightest of the three DMs and the limit can be increased for higher $Z'$ mass). 
The neutralino DM constitutes only 1$\%$-20$\%$ of the total DM density making Eqn. \ref {eqn:2-comp} a good benchmark. 
Note that the scan yields a pre-dominantly wino, but with some Higgsino component in it.  

\begin{figure}[thb]
$$
\includegraphics[height=7cm]{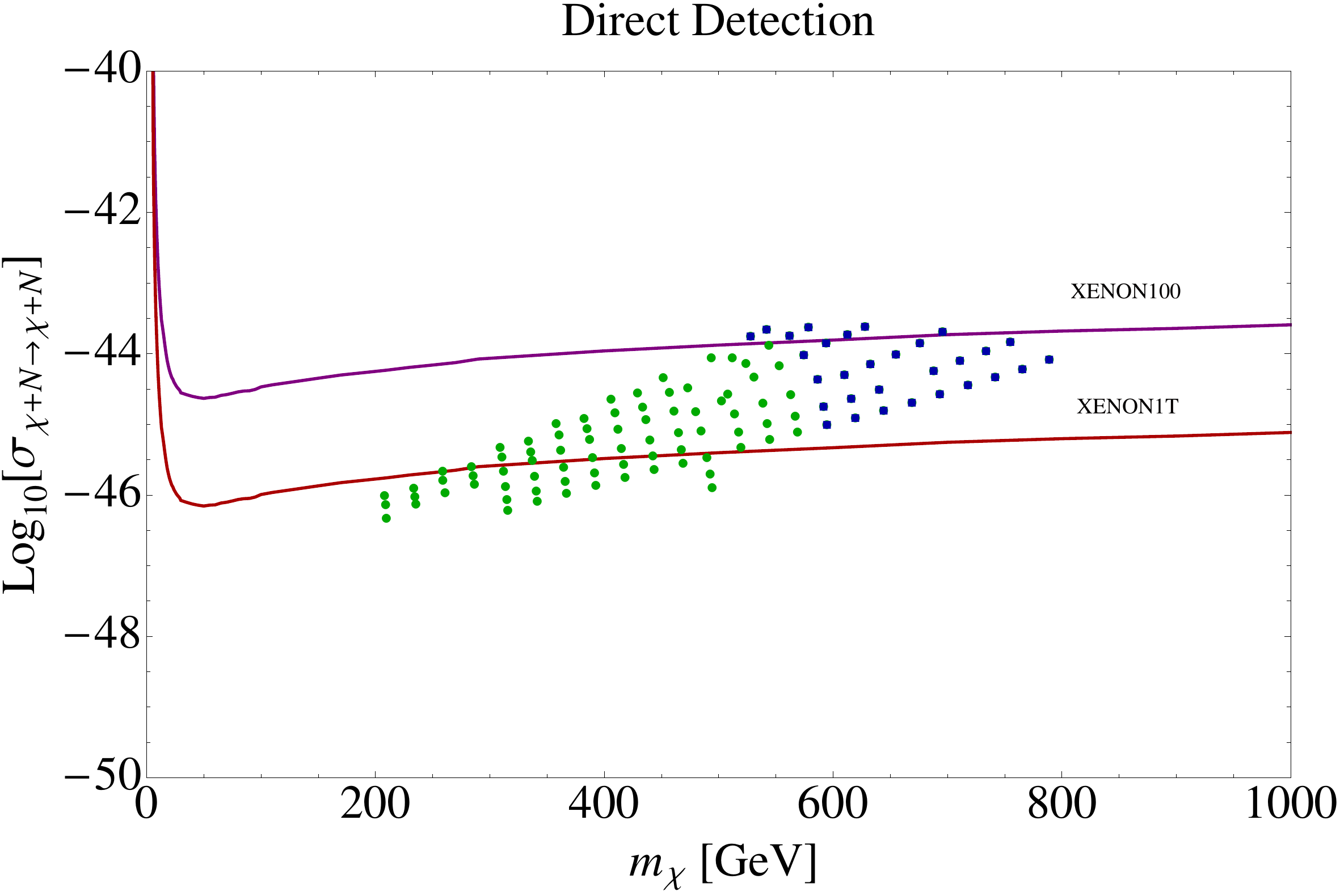}
$$
\caption{Direct detection constraint on neutralino DM mass when it is predominantly wino. The upper curve is for XENON100 and the lower one is for XENON1T. Points 
in blue have relic abundance more than 10$\%$ and those in green have less. Scanned parameter space ranges: $M_1$: (800-1200) GeV, 
$M_2$: (200-775) GeV,  $\mu$: (600-1000) GeV with $\mu, M_1 > M_2$.}
\label{DDMSSM}
\end{figure}

We use MicrOMEGAs \cite{MicrOmega} to evaluate relic abundance and direct detection cross-sections for neutralino DM 
which mimics MSSM in the parameter space mentioned above. 
The direct detection cross-section for neutralino goes through t-channel processes as in Fig. \ref{directChi}. The squark contribution is negligible as they are 
heavy $\simeq$ 2 TeV. Also, for pure wino, there is no Higgs channel and 
the $Z$-channel contributes more to spin- dependent cross-section. Hence, having some Higgsino fraction in the neutralino enhance direct detection. 
in Fig. \ref{DDMSSM}, we see that the neutralino can be accessible to direct detection experiments in near future with 
points close to XENON100 and XENON1T limit. Points in blue have relic abundance contribution with
more than 10$\%$ and they have a early detection possibility while points in green have relic density less than 10$\%$ 
and direct detection for them may be delayed depending on the mass and composition. While higher order 
calculations for direct detection of purely wino DM has been studied \cite{direct} to boost direct detection,
we are not using them, since we are exploiting a small Higgsino fraction in the neutralino, that increases direct detection 
while having co-annihilations to yield under-abundance. 

The mass range and the wino content in neutralino studied here is consistent with the
indirect detection constraints from Fermi Gamma-Ray space telescope or the High 
Energy Spectroscopic System (ÒH.E.S.S.Ó) \cite{constraints}. 

We also note that the MSSM parameter space scan performed here, 
doesn't correspond to a specific high-scale SUSY breaking pattern. So, the bounds on the 
chargino or neutralino masses obtained from LHC \cite{pdg}, which mostly assumes some specific 
high-scale pattern like minimal Supergravity (mSUGRA) \cite{mSUGRA}, are not applicable here.  

\begin{figure}[thb]
$$
\includegraphics[height=7cm]{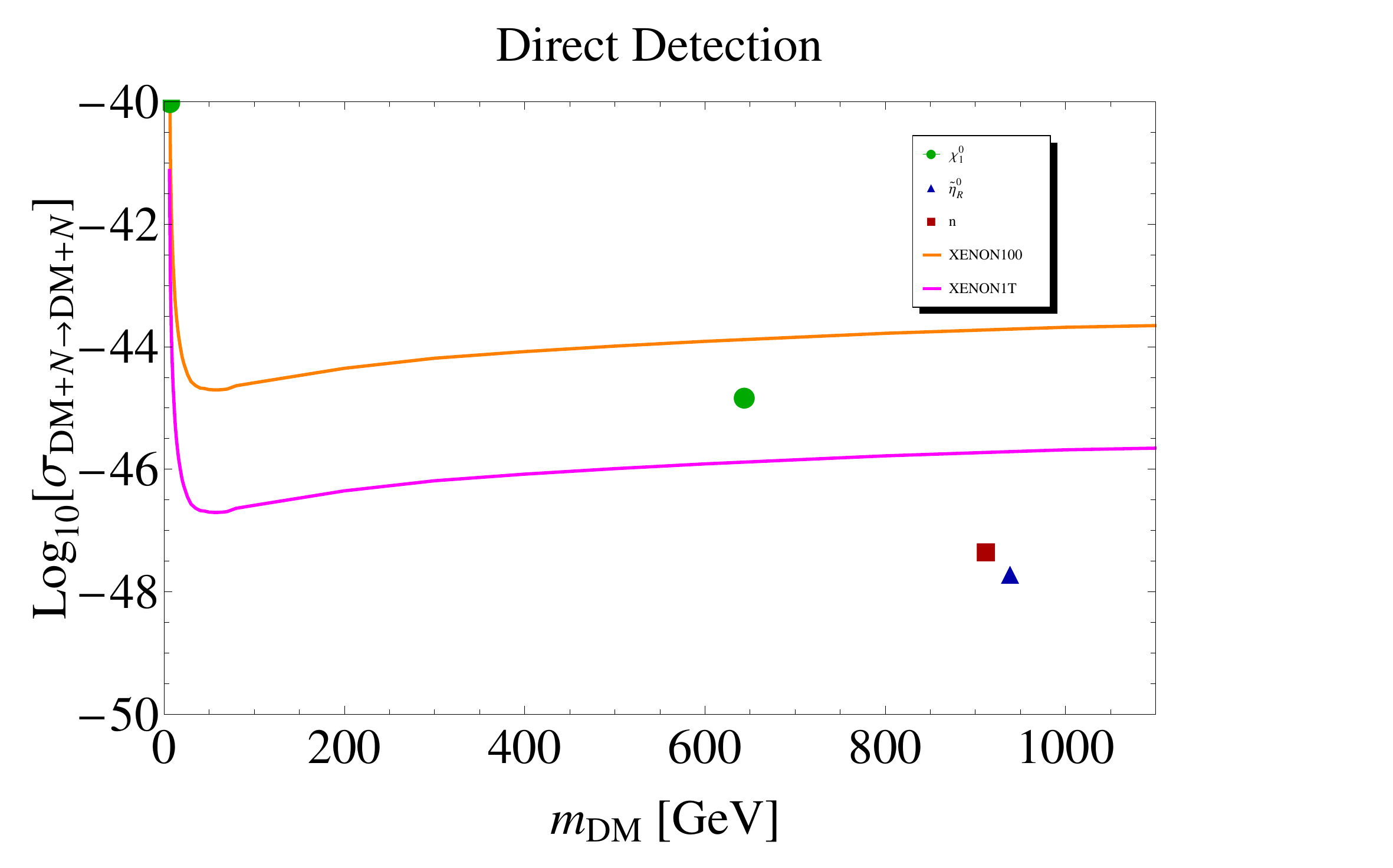}
$$
\caption{A sample point in the three component dark matter parameter space allowed by WMAP is 
plotted with respect to XENON100 and XENON1T limit. $\tilde{\chi}^0_1$ (644 GeV) in green, 
$n$ (912 GeV) in red and $\tilde{\eta}^0_{R}$ (939 GeV) in blue 
constitutes 10.7$\%$, 62.5$\%$ and 26.8$\%$ of total DM density respectively.
}
\label{fig:sample}
\end{figure}
In Fig. \ref{fig:sample}, we show a sample point in the three component dark matter parameter space 
allowed by relic abundance with respect to XENON100 and XENON1T direct detection limit. 
In this point  $\tilde{\chi}^0_1$ (644 GeV) in green, $n$ (912 GeV) in red and $\tilde{\eta}^0_{R}$ (939 GeV) in blue 
constitutes 10.7$\%$, 62.5$\%$ and 26.8$\%$ of total DM density respectively.

\section{Summary and Conclusions}
In extended LR SUSY model three DM components can co-exist together: the lightest neutralino $\tilde{\chi}^0_1$, 
the lightest scotino $n$, and the exotic $\tilde{\eta}^0_{R}$ Higgsino. We show that in the limit of wino dominated 
$\tilde{\chi}^0_1$, thanks to the co-annihilation with chargino to yield under-abundance, 
the other two components contribute heavily to relic abundance, 
with masses $m_n$ and $m_{\eta}$ around 1 TeV that corresponds to the 
resonance annihilation with $m_{DM}\simeq M_Z'/2$. We found a bound on $Z'$ from LHC to be at 2.045 TeV.
With this value of $M_Z'$, the direct detection cross-section for $n$ and $\tilde{\eta}^0_{R}$ is calculated to lie between 
$10^{-47}-10^{-49} $ (cm$^2$)(depending on the fraction in which it contributes to total DM density). This is at least 
an order of magnitude smaller than XENON1T detection limits. Nevertheless, in such a multicomponent set up, 
a large wino dominated neutralino region becomes allowed without much complications while still obeying the 
existing limits and constraints; with appropriate parameters, $\tilde{\chi}^0_1$ does lie within the direct detection limits.

It is worthy to mention that the situation studied in this article is a simplification in the thermal history of
 three component DM set-up. Interaction of DM components (between $n$ and $\tilde{\eta}_R^0$, 
which have been neglected given the specific mass hierarchy), can make the general situation more 
complicated and one needs to solve the coupled Boltzman equations corresponding to $n$, ${\tilde{\eta}_R^0}$ and $\tilde{\chi}_1^0$ to 
study the exact decoupling of each DM component depending on their relative masses and coupling strength. 

The rich particle spectrum of this model with the right handed sector, 
makes it very likely to have interesting collider signatures at LHC by producing these new excitations. 
They also open up new decay channels that may alter the final state event rates in the lepton or jet-rich final 
states with missing energy. This can serve as a distinctive feature of this model from MSSM and change the 
bounds on sparticle masses at LHC. We plan to elaborate on this in a future publication.  



{ \bf Acknowledgements}:

This work is supported in part by the U.~S.~Department of Energy under Grant 
No. DE-FG03-94ER40837. The work of SB is supported by U.S Department of Energy under 
Grant No. DE-SC0008541. The work of DW is supported by UC MEXUS-CONACYT under 
Grant Reference No. 200406-303746.

\section*{Appendix}

Using the basis 
\begin{eqnarray*}
(\tilde{\Psi}^{0})^T=\lbrace\tilde{B},\tilde{W}_{L},\tilde{\phi}_{L1},\tilde{\phi}_{L2},\tilde{W}_{R},\tilde{\delta}_{11},\tilde{\delta}_{22},\tilde{\phi}_{R1},\tilde{\phi}_{R2}\rbrace
\end{eqnarray*}The neutralino mass matrix in our model is :
\begin{equation}
M_{\chi^0}=\left(\begin{array}{cccc|ccccc}
M_{B} & 0 & -\frac{g_{1}v_{L1}}{\sqrt{2}} & \frac{g_{1}v_{L2}}{\sqrt{2}} & 0 & 0 & 0 & -\frac{g_{1}v_{R1}}{\sqrt{2}} & \frac{g_{1}v_{R2}}{\sqrt{2}}\\
0 & M_{L} & \frac{g_{L}v_{L1}}{\sqrt{2}} & -\frac{g_{L}v_{L2}}{\sqrt{2}} & 0 & -\frac{g_{L}u_{1}}{\sqrt{2}} & \frac{g_{L}u_{4}}{\sqrt{2}} & 0 & 0\\
-\frac{g_{1}v_{L1}}{\sqrt{2}} & \frac{g_{L}v_{L1}}{\sqrt{2}} & 0 & -\mu_{L} & 0 & 0 & \frac{f_{1}v_{R2}}{2} & 0 & \frac{f_{1}u_{4}}{2}\\
\frac{g_{1}v_{L2}}{\sqrt{2}} & -\frac{g_{L}v_{L2}}{\sqrt{2}} & -\mu_{L} & 0 & 0 & \frac{f_{2}v_{R1}}{2} & 0 & \frac{f_{2}u_{1}}{2} & 0\\
\hline 0 & 0 & 0 & 0 & M_{R} & -\frac{g_{R}u_{1}}{\sqrt{2}} & \frac{g_{R}u_{4}}{\sqrt{2}} & \frac{g_{R}v_{R1}}{\sqrt{2}} & -\frac{g_{R}v_{R2}}{\sqrt{2}}\\
0 & -\frac{g_{L}u_{1}}{\sqrt{2}} & 0 & \frac{f_{2}v_{R1}}{2} & -\frac{g_{R}u_{1}}{\sqrt{2}} & 0 & -\mu_{\Delta} & \frac{f_{2}v_{L2}}{2} & 0\\
0 & \frac{g_{R}u_{4}}{\sqrt{2}} & \frac{f_{1}v_{R2}}{2} & 0 & \frac{g_{R}u_{4}}{\sqrt{2}} & -\mu_{\Delta} & 0 & 0 & \frac{f_{1}v_{L1}}{2}\\
-\frac{g_{1}v_{R1}}{\sqrt{2}} & 0 & 0 & \frac{f_{2}u_{1}}{2} & \frac{g_{R}v_{R1}}{\sqrt{2}} & \frac{f_{2}v_{L2}}{2} & 0 & 0 & -\mu_{R}\\
\frac{g_{1}v_{R2}}{\sqrt{2}} & 0 & \frac{f_{1}u_{4}}{2} & 0 & -\frac{g_{R}v_{R2}}{\sqrt{2}} & 0 & \frac{f_{1}v_{L1}}{2} & -\mu_{R} & 0
\end{array}\right)
\label{NeutralinoMM1}
\end{equation}

As mentioned in Section 4.3, this is also not the full matrix, but some elements are already decoupled form this matrix. It is important to note, 
$\tilde{B}$ is not the MSSM $U(1)_Y$ bino, instead is a $SU(2)_R\times U(1)_Y$. But a linear combination 
of $\tilde{B}$ with $\tilde{W_R}$ in the limit $M_B \simeq M_R$ makes it a MSSM bino. 

Defining the ratios $R_W=v_L/v_{SM}=\sqrt{\frac{1-2s_{W}^{2}}{1-s_{W}^{2}}}=0.837$ and $\tan\beta_{L}=v_{L2}/v_{L1}$, we can rewrite the 4$\times$4
upper left matrix in (\ref{NeutralinoMM1}), with the basis $\left\{ \tilde{B},\tilde{W},\tilde{\phi}_{L1,}\tilde{\phi}_{L2}\right\} $,
(where $\tilde{W}=\frac{\tilde{W}_{L}}{R_W}$)  as 
 
\begin{equation}
M_{\chi^0_{MSSM}}=\left(\begin{array}{cccc}
M_{B} & 0 & -M_{Z}*s_{W}*c_{\beta L} & M_{Z}*s_{W}*s_{\beta L}\\
0 & \frac{M_{L}}{R_W^{2}} & M_{Z}*c_{W}*c_{\beta L} & -M_{Z}*c_{W}*s_{\beta L}\\
-M_{Z}*s_{W}*c_{\beta L} & M_{Z}*c_{W}*c_{\beta L} & 0 & -\mu_{L}\\
M_{Z}*s_{W}*s_{\beta L} & -M_{Z}*c_{W}*s_{\beta L} & -\mu_{L} & 0
\end{array}\right)
\label{NeutralinoMM2}
\end{equation}
 
This is exactly the MSSM neutralino mass matrix,
where $M_{B}=M_{1},\mu_{L}=\mu,\beta_{L}=\beta,$and $\frac{M_{L}}{R_W^{2}}\simeq1.43M_{L}=M_{2}$.

The rest of the Higgsinos in the basis, $\left(\tilde{\Psi_{2}}^{0}\right)^{T}=\lbrace\tilde{\eta_{L1}^{0}},\tilde{\eta_{L2}^{0}},\tilde{s_{3}^{0}}\rbrace$
 , $\left(\tilde{\Psi_{3}}^{0}\right)^{T}=\lbrace\tilde{\eta_{R1}^{0}},\tilde{\eta_{R2}^{0}}\rbrace$
  and $\left(\tilde{\Psi_{4}}^{0}\right)^{T}=\lbrace\tilde{\delta_{12}^{0}},\tilde{\delta_{21}^{0}}\rbrace$
  are 
\begin{eqnarray}
M_{H2}=\begin{pmatrix}0 & -\mu_{L2} & f_{10}v_{L2}\\
-\mu_{L2} & 0 & f_{9}v_{L1}\\
f_{10}v_{L2} & f_{9}v_{L1} & -\mu_{s3}
\end{pmatrix}, M_{H3}=\begin{pmatrix}0 & -\mu_{R2}\\
-\mu_{R2} & 0
\end{pmatrix}, M_{H4}=\begin{pmatrix}0 & -\mu_{\Delta}\\
-\mu_{\Delta} & 0
\end{pmatrix}
\end{eqnarray}

In the following basis:
\begin{eqnarray*}
(\Psi_{1}^{+})^{T}=\lbrace\imath\tilde{W}_{L}^{+},\imath\tilde{W}_{R}^{+},\tilde{\phi_{L2}^{+}},\tilde{\delta_{22}^{+}},\tilde{\phi_{R2}^{+}},\tilde{\delta_{12}^{+}}\rbrace\\
(\Psi_{1}^{-})^{T}=\lbrace\imath\tilde{W}_{L}^{-},\imath\tilde{W}_{R}^{-},\tilde{\phi_{L1}^{-}},\tilde{\delta_{11}^{-}},\tilde{\phi_{R1}^{-}},\tilde{\delta_{21}^{-}}\rbrace\\
(\Psi_{2}^{+})^{T}=\lbrace\tilde{\eta}_{L2}^{+},\tilde{\eta}_{R2}^{+},\zeta_{2}^{+}\rbrace\\
(\Psi_{2}^{-})^{T}=\lbrace\tilde{\eta}_{L1}^{-},\tilde{\eta}_{R1}^{-},\zeta_{1}^{-}\rbrace
\end{eqnarray*}

The chargino mass matrices are:
\begin{eqnarray}
M_{1\chi^{\pm}}=\left(\begin{array}{cccccc}
M_{L} & 0 & \frac{g_{L}v_{L2}}{2} & \frac{g_{L}u_{4}}{2} & 0 & 0\\
0 & M_{R} & 0 & \frac{g_{R}u_{4}}{2} & \frac{g_{R}v_{R2}}{2} & 0\\
\frac{g_{L}v_{L1}}{2} & 0 & \mu_{L} & -f_{1}v_{R2} & 0 & 0\\
\frac{g_{L}u_{1}}{2} & \frac{g_{R}u_{1}}{2} & -f_{2}v_{R1} & \mu_{\Delta} & 0 & 0\\
0 & \frac{g_{R}v_{R1}}{2} & 0 & 0 & \mu_{R} & -f_{2}v_{L2}\\
0 & 0 & 0 & 0 & -f_{1}v_{L1} & \mu_{\Delta}
\end{array}\right)\label{charginomatrix}
\end{eqnarray}

\begin{equation}
M_{2\chi^{\pm}}=\begin{pmatrix}\mu_{L2} & f_{3}u_{1} & f_{5}v_{L1}\\
f_{4}u_{4} & \mu_{R2} & f_{6}v_{R1}\\
-f_{7}v_{L2} & -f_{8}v_{R2} & \mu_{s12}
\end{pmatrix}\label{charginomatrix2}
\end{equation}

\newpage
\bibliographystyle{unsrt}

\end{document}